\documentclass[aps,pra,reprint,groupedaddress]{revtex4-2}

\usepackage[utf8]{inputenc}
\usepackage{amsmath}
\usepackage{amsfonts}
\usepackage{amssymb}
\usepackage{braket}
\usepackage{bbm}
\usepackage{graphicx}
\usepackage{tikz}
\usepackage{changepage}
\usepackage{multirow}
\usepackage{mathtools}
\usepackage{notes2bib}

\def \l( {\left(}
\def \r) {\right)}

\def \b {\hat{b}}
\def \bplus {\hat{b}^\dagger}
\def \Z {\mathcal{Z}}

\def \C {\mathcal{C}}

\def \n {\hat{n}}
\def \e {\epsilon}
\def \O {\mathcal{O}}

\def \L {\mathcal{L}}
\def \R {\mathcal{R}}
\def \ULR {U_{\L\R}}
\def \ttilde {\tilde{t}}

\def \mutilde {\tilde{\mu}}
\def \Vtilde {\tilde{V}}
\def \Udtilde {\tilde{U}_d}
\def \ULRtilde {\tilde{U}_{\L\R}}

\def \H {\mathcal{\hat{H}}}
\def \Hon {\mathcal{\hat{H}}_{\text{on}}}
\def \Hoff {\mathcal{\hat{H}}_{\text{off}}}

\begin{document}

% Use the \preprint command to place your local institutional report
% number in the upper righthand corner of the title page in preprint mode.
% Multiple \preprint commands are allowed.
% Use the 'preprintnumbers' class option to override journal defaults
% to display numbers if necessary
%\preprint{}

%Title of paper
\title{The double-well Bose Hubbard model with nearest-neighbor and cavity-mediated long-range interactions}

% repeat the \author .. \affiliation  etc. as needed
% \email, \thanks, \homepage, \altaffiliation all apply to the current
% author. Explanatory text should go in the []'s, actual e-mail
% address or url should go in the {}'s for \email and \homepage.
% Please use the appropriate macro foreach each type of information

% \affiliation command applies to all authors since the last
% \affiliation command. The \affiliation command should follow the
% other information
% \affiliation can be followed by \email, \homepage, \thanks as well.
\author{Johannes Sicks}
\email[]{johannes@lusi.uni-sb.de}
%\homepage[]{Your web page}
%\thanks{}
%\altaffiliation{}
\author{Heiko Rieger}
\affiliation{Theoretical Physics, Saarland University, Campus E2.6, 66123 Saarbr{\"u}cken, Germany}

%Collaboration name if desired (requires use of superscriptaddress
%option in \documentclass). \noaffiliation is required (may also be
%used with the \author command).
%\collaboration can be followed by \email, \homepage, \thanks as well.
%\collaboration{}
%\noaffiliation

\date{\today}

\begin{abstract}
We consider a one-dimensional Bose-Hubbard model (BHM) with on-site double-well potentials and study the effect of nearest-neighbor repulsion and cavity-mediated long-range interactions by calculating the ground-state phase diagrams with quantum Monte-Carlo simulations. We show that when the intra-well repulsion is as strong as the on-site repulsion a dimerized Mott insulator phase appears at the tip of the dimerized Density Wave phase for a density of one particle per double well. Furthermore, we find a dimerized Haldane insulator phase in the double-well BHM with nearest-neighbor interaction, which is identical to a dimerized BHM with repulsive interactions up to the third neighbor.
\end{abstract}

% insert suggested keywords - APS authors don't need to do this
%\keywords{}

%\maketitle must follow title, authors, abstract, and keywords
\maketitle

% body of paper here - Use proper section commands
% References should be done using the \cite, \ref, and \label commands
\section{Introduction}
\label{intro}

Since its introduction, the Bose-Hubbard-Model (BHM) \cite{Fisher1989} has been a focus of research. In the simplest form, where tunneling between neighboring sites and a repulsive on-site interaction exist, the ground state phase diagram is characterized by two phases. For commensurate fillings and large on-site repulsions compared to the bosonic tunneling, a Mott insulator (MI) phase appears, while for incommensurate fillings or weak on-site repulsion, the Superfluid (SF) phase exists.\\
Jaksch \textit{et al.} showed in \cite{Jaksch1998}, that the dynamics of ultracold bosons, which are contained by an optical lattice, realize a BHM. This was experimentally shown by Greiner \textit{et al.} \cite{Greiner2002} and led consequentially to a broad study of experiments of ultracold bosons in optical lattices \cite{Landig2016,Anderlini2006,Anderlini2007,Atala2013,Baumann2010}.
\\
The universality class of the SF-MI phase transition in the BHM is generally of mean-field type, except for the multicritical point, where particle-hole symmetry holds \cite{Fisher1989}. At this point, the universality class changes to the type of the ($d$+1) dimensional XY model \cite{Lieb1961,Kogut1979,Ma1986}, where the two dimensional XY model \cite{Tobochnik1979,Mattis1984} has a topological Kosterlitz-Thouless phase transition \cite{Kosterlitz1973,Kosterlitz1974}. 
Furthermore, when the occupation per site is restricted to 0, 1 and 2 bosons per site, the one-dimensional BHM can be described by a quantum spin 1 chain, which features a gapped Haldane insulator phase, characterized by a nonlocal order parameter \cite{DallaTorre2006,Haldane1983,Haldane1983a,denNijs1989,Berg2008}.\\
The quantum critical phenomena of the BHM were studied extensively with quantum Monte-Carlo (QMC) methods, like the path-integral \cite{Pollock1987,Ceperley1995}, world-line \cite{Batrouni1992,Batrouni1995} and worm-algorithm QMC \cite{Prokofev1998,Prokofev1998a}, of which we use the latter in this work. Also approximate methods were used, like the mean-field theory \cite{Fisher1989,Oosten2001} and density matrix renormalization group method \cite{Kuehner1998}.\\
There are many different modifications and extensions to the BHM, originating from the addition of interactions or changes to the confining potentials. Possible interactions are nearest-neighbor interaction \cite{Kuehner1998,Batrouni2013,Ohgoe2012,Sengupta2005,Batrouni2006,Iskin2011,Rossini2012,Kawaki2017} (also referred to as extended BHM), next-nearest-neighbor interaction \cite{Schmid2004,DallaTorre2006} and hopping \cite{Chen2008}, cavity-mediated long-range interaction \cite{Landig2016,Dogra2016,Flottat2017,Hruby2018} and combinations of nearest-neighbor and long-range interaction \cite{Bogner2019,Sicks2020}.
Changes to the confining potentials include for instance disordered potentials \cite{Gurarie2009,Niederle2016} and double-well potentials \cite{SebbyStrabley2006,Foelling2007,Trotzky2008,Barmettler2008,Yin2015,Volosniev2015,Mondal2019}.\\
Superimposing two optical lattices with different wavelengths form a so-called superlattice, like the double-well lattice in which each site consists of a double-well potential \cite{SebbyStrabley2006,Peil2003}. With the help of double-well lattices, quantum information processes can be studied \cite{Brennen1999,Yang2020}, as it allows for example to manipulate atoms individually \cite{Lee2007} or study the many-body dynamics and entanglement of a double-well chain \cite{Barmettler2008}. Furthermore, the hard-core bosonic double-well BHM is the bosonic counterpart of the Su-Schrieffer-Heeger model \cite{Su1979} for free fermions, which possesses a nontrivial topological insulator phase. For the hard-core double-well BHM, this topological phase was shown as well \cite{Mondal2019,Grusdt2013} and the ground state properties were studied recently \cite{Hayashi2022}.\\
In this paper, we study the ground state phase diagram of the one-dimensional extended double-well BHM with cavity-mediated long-range interaction. In this model, each lattice site consists of one double-well potential, which are aligned in a chain. We consider nearest-neighbor interaction between the sites, therefore bosons in both wells of the double-well on one site feel the interactions between all wells of the neighboring double-well sites. Also, the intra-well repulsion between the two wells of each double-well is taken into account.\\
Its parameter space includes a one-dimensional extended single-well BHM with long-range interaction, where additional Density Wave (DW), Supersolid (SS) and Haldane insulator (HI) phases appear \cite{Sicks2020} and a one-dimensional dimerized BHM, where intra-well repulsion and hopping strength alternate between every other site \cite{Atala2013,Hayashi2022,Kundu2009,DiLiberto2017,Sugimoto2019,Azcona2021} and bond-ordered phases appear \cite{Nakamura1999}.\\
We are interested in the similarities and differences in the ground state behavior between the double-well BHM and the single-well and dimerized BHM. While the effects of cavity-mediated long-range interaction on the single-well BHM were studied \cite{Landig2016,Dogra2016}, its effects on a double-well lattice chain remain unknown. Also the interaction between neighboring double-wells - which corresponds to an interaction range up to the third neighbor in the dimerized chain - has not been considered yet.\\
~\\
The paper is organized as follows: In Section \ref{sec:Model_and_Mean-Field_Considerations}, the Hamiltonian of the one-dimensional extended double-well BHM with cavity-mediated long-range interaction is defined and the order parameters are introduced. Then, the analytically solvable ground states without hopping terms are discussed. Section \ref{sec:Simulation_Results} contains the QMC worm-algorithm results for the ground states of the standard and extended double-well BHM. First, we examine the standard double-well BHM, before giving results for the extended double-well BHM and double-well BHM with cavity-mediated long-range interaction. The conclusion is given in Section \ref{sec:Conclusion}.

\section{Model}
\label{sec:Model_and_Mean-Field_Considerations}

\subsection{Hamiltonian of the Double-Well BHM}
\label{sec:Hamiltonian_of_the_Double-Well_BHM}

\begin{figure}[t]\centering
\includegraphics[width=\columnwidth]{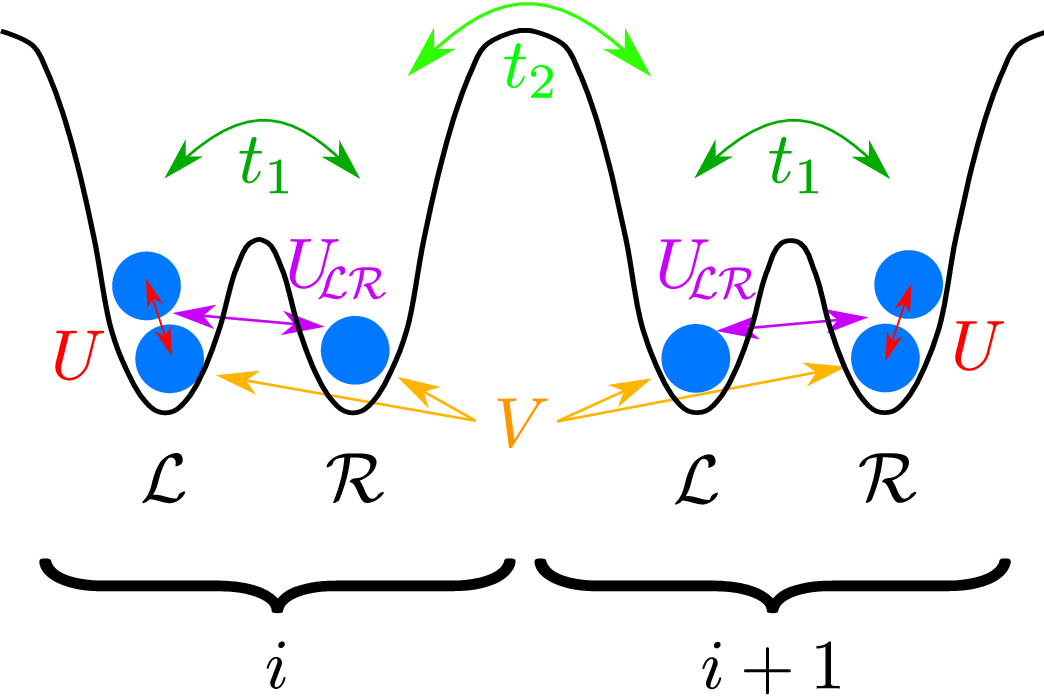}%
\caption{Sketch of two sites of the double-well Bose-Hubbard model according to the interactions in Hamiltonian (\ref{eq:H}) without the $\H_{U_d}$ term. Each site consists of a double-well with a left ($\L$) and a right ($\R$) well. In each well, bosons experience an on-site repulsion $U$, when two or more bosons are present. Intra-well tunneling $t_1$ is possible between left and right well of the same site and inter-well tunneling $t_2$ between left and right well of adjacent sites. Intra-well repulsion $U_{\L\R}$ is present between left and right well of the same site and the nearest-neighbor interaction $V$ interacts between all wells of adjacent sites.}
\label{fig:double-well}
\end{figure}
We state the one-dimensional extended double-well Bose Hubbard Model (BHM) Hamiltonian with cavity-mediated long-range interaction. In Fig. \ref{fig:double-well} we depict all interactions of the Hamiltonian except for the cavity-mediated long-range interaction. Each site position contains one double-well and is labeled by the index ${i \in 1,\dots,L}$, while the left and right well of the respective double-well is given by $\sigma \in \lbrace \L,\R \rbrace$. Here, $L$ is the length of the chain. We use periodic boundary conditions, thus ${L+1,\sigma} \equiv {1,\sigma}$.\\
The Hamilton operator for the one-dimensional extended double-well BHM with cavity-mediated long-range interaction reads 
\begin{equation}
\H = \H_t + \H_U + \H_{U_{\L\R}} + \H_V + \H_{\mu} + \H_{U_d} ,
\label{eq:H}
\end{equation}
where the particular terms have the following form.\\
$\H_t$ depicts the hopping terms
\begin{equation}
\H_{t} =  - t_1 \sum_i  \left( \b^{\dagger}_{i,\L}\b_{i,\R} + h.c. \right) - t_2 \sum_i  \left( \b^{\dagger}_{i,\R}\b_{i+1,\L} + h.c. \right) ,
\label{eq:H_t}
\end{equation}
where $t_1$ is the intra-well hopping parameter between left and right wells of a single double-well on each site and $t_2$ the inter-well hopping parameter between adjacent left and right wells of double-wells next to each other.
\begin{equation}
\H_{U} = \dfrac{U}{2} \sum_{\substack{i \\ \sigma=\L,\R}} \n_{i,\sigma} (\n_{i,\sigma} - 1)
\label{eq:H_U}
\end{equation}
is the on-site interaction on each site, separated between the two wells and
\begin{equation}
\H_{U_{\L\R}} = U_{\L\R} \sum_i \n_{i_\L} \n_{i,\R}
\label{eq:H_U_LR}
\end{equation}
defines the intra-well repulsion between bosons located in different wells on the same site. The repulsive interaction between neighboring sites is given by
\begin{equation}
\H_{V} = V \sum_{\substack{i \\ \sigma,\sigma'=\L,\R}} \n_{i,\sigma} \n_{i+1,\sigma'} \equiv V \sum_{i} \n_i \n_{i+1}.
\label{eq:H_V}
\end{equation}
\begin{figure*}[t]\centering
\includegraphics[width=1.5\columnwidth]{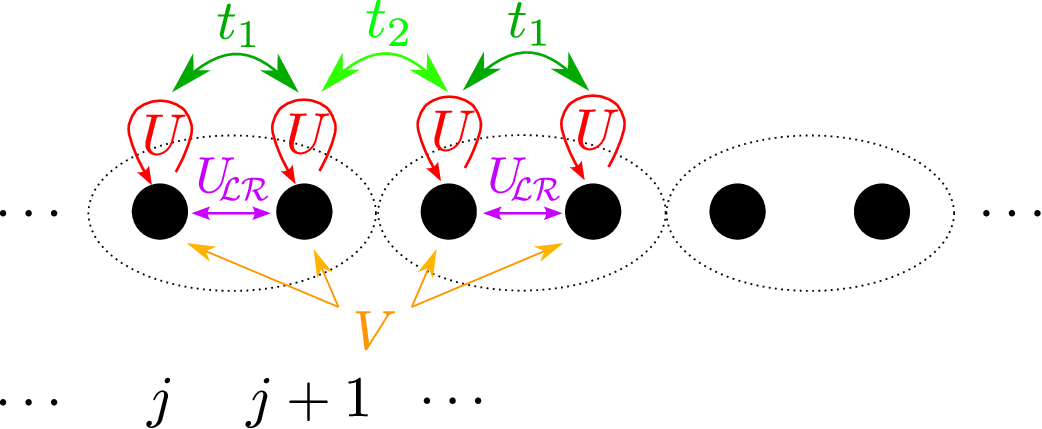}%
\caption{Sketch of the dimerized chain described by Eq. (\ref{eq:H}). The $\H_{U_d}$ term is not depicted. The dotted circles around two sites is a guide to the eye and correspond to the double-wells of Fig. \ref{fig:double-well}. The indices $i,\L$ ($i,\R$) match $j$ ($j+1$) and result in a dimerized chain length twice as long as the double-well chain.}
\label{fig:dimerized_chain}
\end{figure*}\\
Here we assume, that the spatial distance between neighboring sites is much larger than the distance between left and right well of the double-well on one site and therefore that $V$ is independent of the well index $\sigma$.\\
To abbreviate the notation, we define $\n_i = \n_{i,\L} + \n_{i,\R}$ and can omit most of the $\sigma$-indices in the Hamiltonians. The chemical potential term is
\begin{equation}
\H_{\mu} = - \mu \sum_{\substack{i \\ \sigma=\L,\R}} \n_{i,\sigma} \equiv - \mu \sum_{i} \n_i
\label{eq:H_mu}
\end{equation}
with the total boson number operator $\hat{N} = \sum_{i} \n_i$.\\
The last Hamiltonian
\begin{align}
\H_{U_d} = &- \dfrac{U_d}{L} \left( \sum_{\substack{i~\text{even} \\ \sigma=\L,\R}} \n_{i,\sigma} - \sum_{\substack{i~\text{odd} \\ \sigma=\L,\R}} \n_{i,\sigma} \right)^2 \nonumber \\
=&- \dfrac{U_d}{L} \left( \sum_{i~\text{even}} \n_{i} - \sum_{i~\text{odd}} \n_{i} \right)^2
\label{eq:H_U_d}
\end{align}
represents the cavity-mediated long-range interaction between even and odd chain sites.\\
Hamiltonian (\ref{eq:H}) is identical to a dimerized chain with
\begin{equation}
\H_{t_{\text{dim}}} =  - t \sum_{j}\l ( 1 + (-1)^{j+1}\delta \r)  \left( \b^{\dagger}_{j}\b_{j+1} + h.c. \right) ,
\label{eq:H_t_dim}
\end{equation}
where $t = (t_1+t_2)/2$ is the mean hopping strength and $\delta = (t_1-t_2)/(t_1+t_2)$ the bond dimerization. Likewise, the intra-well repulsion can be understood as a dimerized nearest-neighbor interaction
\begin{equation}
\H_{\ULR,~\text{dim}} =  - \dfrac{\ULR}{2} \sum_{j}\l( 1 + (-1)^{j+1} \r)  \n_j \n_{j+1} .
\label{eq:H_ULR_dim}
\end{equation}
The index $j$ is hereby the combination of the $i,\sigma$ notation into one index, where $j \equiv i,\L$ and $j+1 \equiv i,\R$. Therefore, the chain length is doubled.\\
Fig. \ref{fig:dimerized_chain} represents Hamiltonian (\ref{eq:H}) in form of a dimerized chain BHM, where the depicted interactions correspond to Fig. \ref{fig:double-well}. The cavity-mediated long-range interaction is not shown. Each double-well is hereby equal to a pair of sites in the dimerized chain, highlighted by the dotted circle around the pair. Therefore, even and odd sites of the double-well BHM are even and odd \emph{pair of sites} of the dimerized chain, which is important to note for the cavity-mediated long-range interaction and the definition of the notation of the phases used in this work.\\
The intra-well repulsion acts as a dimerized interaction itself and can be regarded for instance as an inter-chain nearest-neighbor interaction of a two-leg ladder model \cite{Singh2014}, however, the alignment of the sites is different between this model and the double-well BH chain used in this work.

\subsection{Simulation Method and Order Parameters}

\begin{figure*}[t]\centering
\includegraphics[width=1.5\columnwidth]{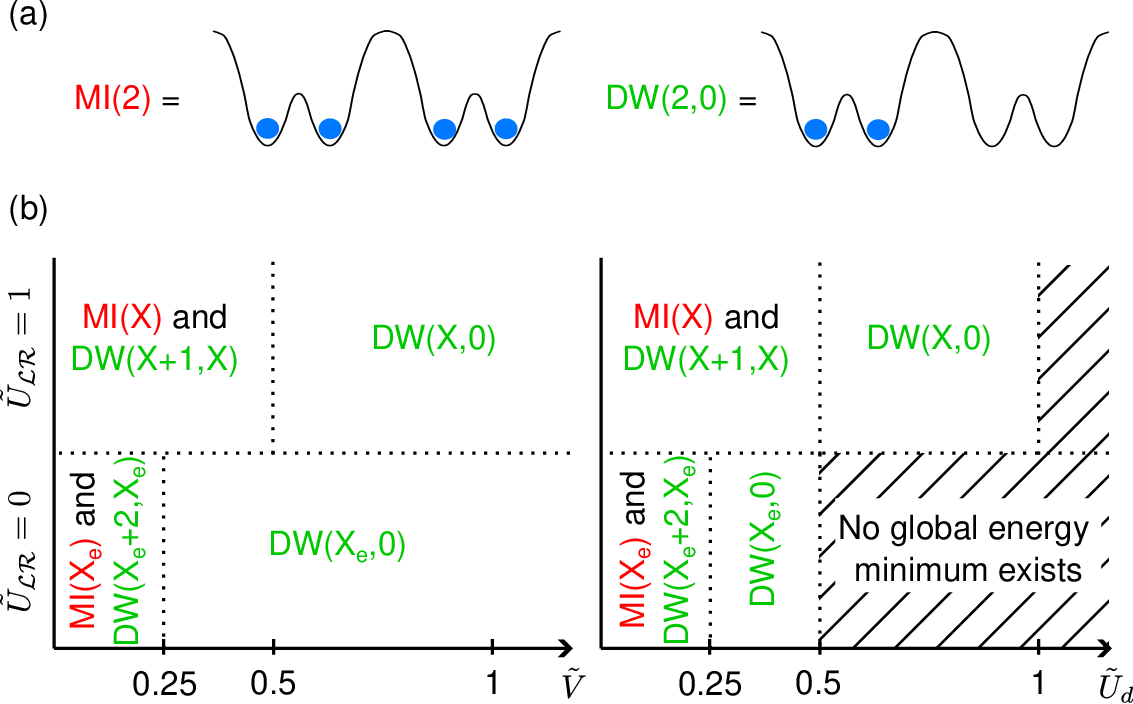}%
\caption{(a) Depiction of the MI(2) and DW(2,0) phases. In the former, two bosons are located on every site, while in the latter, only every other site is occupied by two bosons. (b) Ground state sketch of Hamiltonian (\ref{eq:H}) without hopping terms ($t_1=t_2=0$) for $\Vtilde>0,~\Udtilde=0$ (left) and  $\Vtilde=0,~\Udtilde>0$ (right), in the two cases $\ULRtilde=0$ and $\ULRtilde=1$. $X$ is an integer number, while $X_e$ represents only even integer numbers. When $\Vtilde=0,~\Udtilde=0$, DW phases vanish and only MI phases persist.}
\label{fig:ground_states_sketch}       % Give a unique label
\end{figure*}

We use the exact quantum Monte-Carlo (QMC) worm-algorithm \cite{Prokofev1998,Prokofev1998a} to obtain the phase diagrams. This method operates in the grand-canonical ensemble, thus, the boson number is not fixed. We consider chain lengths up to $L=64$, where each site consists of one double-well potential. We elaborate the QMC-WA further in Appendix A.\\
From the QMC-WA simulations we obtain the boson density 
\begin{equation}
\rho = \frac{1}{L}\sum_i \langle \n_i \rangle
\end{equation}
and the superfluid density
\begin{equation}
 \rho_s = \frac{\langle W^2 \rangle L}{2t_2\beta},
 \end{equation} with $W$ being the winding number, which is defined as the difference between boson lines crossing the periodic boundary condition in one direction versus the other direction. Furthermore with the density-density correlation $D(r)= \frac{1}{L}\sum_i \langle \n_i \n_{i+r} \rangle$, the structure factor is defined as
\begin{equation}
S(k) = \dfrac{1}{L} \sum_r e^{\imath k r} D(r).
\end{equation}
With these order parameters we are able to distinguish between the Mott insulator (MI), Superfluid (SF), Density Wave (DW) and Supersolid (SS) phases. As shown in single-well BHM with nearest-neighbor interactions, the so called Haldene insulator (HI) phase can emerge at the tip of the DW lobes \cite{DallaTorre2006,Berg2008}, originating from the spin $1$ antiferromagnetic Heisenberg chain \cite{Haldane1983,Haldane1983a}. To determine the HI we must introduce two non-local observables, the string and parity operators,
\begin{align}
\O_s(|i-j|) &= \left\langle \delta\n_i \exp \left\lbrace \imath\pi \sum_{k=i}^{j} \delta\n_k \right\rbrace \delta\n_j  \right\rangle, \\
\O_p(|i-j|) &= \left\langle \exp \left\lbrace \imath\pi \sum_{k=i}^{j} \delta\n_k \right\rbrace  \right\rangle,
\end{align}
where $\delta\n_i = \n_i - \rho$ is the difference between particle number and density. Due to periodic boundary conditions, both observables are evaluated for $|i-j|=L/2$.\\
The upper part in Tab. \ref{tab:transitions} shows the phases, which can be identified in the ground state phase diagram with the help of the above mentioned order parameters. As notation, we use for Mott insulator phases MI(X), where X is the number of bosons in each double-well. For the Density Wave phases we use DW(X,Y), with X being the boson number on even sites and Y the boson number on odd sites (see Fig. \ref{fig:ground_states_sketch} (a)).
\begin{table}[htbp] 
%\begin{ruledtabular}
\centering
\begin{tabular}{l|c c c c c c}
%&&&&&&\\
& $\rho_s$ & $S(\pi)$ & $\O_s\l( \frac{L}{2} \r) $ & $\O_p\l( \frac{L}{2} \r) $ & $\zeta$ & $\Delta$ \\  
%&&&&&&\\
\hline 
SF & $\neq 0$ & $0$ & $0$ & $0$ & $\neq0$ & $\neq0$ \\ %\hline 
SS & $\neq 0$ & $\neq 0$ & $\neq 0$ & $\neq 0$ & $\neq0$ & $\neq0$ \\ %\hline
DW($X_o$,0) & $0$ & $\neq 0$ & $\neq 0$ & $\neq 0$ & $0$ & $1/2$\\ %\hline 
DW($X_e$,0) & $0$ & $\neq 0$ & $\neq 0$ & $\neq 0$ & $0$ & $0$\\
MI($X_o$) & $0$ & $0$ & $0$ & $\neq 0$ & $0$ & $1$ \\ %\hline 
MI($X_e$) & $0$ & $0$ & $0$ & $\neq 0$ & $0$ & $0$ \\
HI(1) & $0$ & $0$ & $\neq 0$ & $0$ & $0$ & $\neq0$ \\ 
\hline
D-DW($X_o$,0) & $0$ & $\neq 0$ & $\neq 0$ & $\neq 0$ & $\neq0$ & $\neq0$\\ %\hline 
D-DW($X_e$,0) & $0$ & $\neq 0$ & $\neq 0$ & $\neq 0$ & $\neq0$ & $\neq0$\\
D-MI($X_o$) & $0$ & $0$ & $0$ & $\neq 0$ & $\neq0$ & $\neq0$ \\ %\hline 
D-MI($X_e$) & $0$ & $0$ & $0$ & $\neq 0$ & $\neq0$ & $\neq0$ \\
D-HI(1) & $0$ & $0$ & $\neq 0$ & $0$ & $\neq0$ & $\neq0$ \\ 
\end{tabular}
\caption{Order parameters for the phases studied in this paper. We differentiate between (dimerized) MI and DW phases with even (e) or odd (o) particle numbers per double-well.}
\label{tab:transitions}
%\end{ruledtabular}
\end{table}\\
To further differentiate the behavior of the double-well dynamics, we introduce the intra-well fluctuation parameter
\begin{equation}
\zeta \propto \langle \b_{i,\L} \bplus_{i,\R} + h.c. \rangle
\end{equation}
as an indicator for the bosonic movement inside a double-well between the left and right well. It is linked to the kinetic energy operator for dimerized models \cite{Kundu2009}. Furthermore we define the well occupation difference
\begin{equation}
\Delta = \dfrac{1}{L}\sum_i \langle | \n_{i,\L} - \n_{i,\R} | \rangle .
\end{equation}
When $\Delta = 0$ the boson distribution inside a double-well is symmetric, meaning that as many particle are present in the left well, as in the right well for every site. If $\Delta > 0 $ the symmetry is (partially) broken, as it happens when intra-well fluctuations become stronger. When $\zeta = 0$ all fluctuations inside the double-wells vanish. For $\zeta>0$ and $\rho_s=0$ the movement inside the double-wells can be compared to the dimerized BHM, where the finite bond dimerization leads to dimerized Mott insulator (D-MI), dimerized Density Wave (D-DW) and dimerized Haldane insulator (D-HI) phases \cite{Kundu2009}. The bottom part of Tab. \ref{tab:transitions} shows the dimerized phases, which are characterized by $\zeta$ and $\Delta$.\\
\\
In the following, we express all parameters in units of the on-site repulsion by the abbreviated form $\mutilde = \mu/U$ and analogous for all other parameters.

\subsection{Ground States without Hopping Terms}
\label{sec:Ground_States_without_Hopping_Terms}
The ground state phase diagram of Hamiltonian (\ref{eq:H}) can be calculated analytically when the hopping terms are neglected, because the number operator is diagonal in the Fock basis. In analogy to the ground state phase diagrams of the single-well BHM and dimerized BHM we expect Mott insulator (MI) phases and Density Wave (DW) phases to appear.\\
We do not differentiate between various boson configurations inside each double-well. This is determined by the ratio of on-site repulsion and intra-well repulsion $ U_{\L\R}/U = \ULRtilde$. If $\ULRtilde < 1$, the symmetric filling of left and right wells on each site is most favorable and in case of odd boson numbers, the last particle can be located in either well, resulting in two equal likely configurations for each double-well site.\\
In the case $\ULRtilde = 1$, on-site repulsion and intra-well repulsion are equally strong. Thus, the energy gain by increasing the boson number in one well by one and the energy gain by increasing the boson number in the neighboring well on one site are the same. As a result, all boson distributions inside the double-well share the same energy and the degeneracy can be determined by combinatorics. To distribute X bosons in two wells, there are $\binom{X+2-1}{X} = X+1$ possible arrangements per site.\\
Fig. \ref{fig:ground_states_sketch} shows the ground state phase diagram of the double-well BHM with cavity-mediated long-range interaction $U_d$ and nearest-neighbor interaction $V$ for $\ULRtilde = 0$ and $\ULRtilde = 1$.\\
Generally, nearest-neighbor and cavity-mediated long-range interaction share a lot of commonalities in their ground state behavior, like on a mean-field level, where they show identical phase diagrams \cite{Dogra2016}. For $\ULRtilde=0$, the occupation of left and right well on each site is symmetric, thus only even particle numbers per site occur. One difference is the behavior, when the interaction becomes sufficient large compared to the on-site repulsion, as for the cavity-mediated long-range interaction there is no global energy minimum and thus no ground state anymore. One can see the reason for this in the energy per site, which the system is gaining via the nearest-neighbor interaction and the long-range interaction
\begin{align}
\epsilon_{V} &= 4VXY , \nonumber \\
\epsilon_{U_d} &= - \dfrac{U_d}{4} \l( X^2 + Y^2 - 2XY \r) ,
\label{eq:energy_per_site}
\end{align}
where $X$ and $Y$ give the particle number on even and odd sites. Apparently, the nearest-neighbor interaction increases the energy, when neighboring sites are occupied. This includes the MI phases, while DW phases remain unaffected. This reverses for the long-range interaction, where the energy is lowered, when a misbalance of particle occupation between even and odd sites is present, as for the DW phases, while the MI phases are unaffected from the long-range interaction. Hence, when the energy lowering from the long-range interaction is more than the energy gain from the on-site repulsion, the global energy function becomes a concave function and no energy minimum exists anymore.\\
~\\
We can also explain why the transition points for $\ULRtilde=0$ are halved compared to $\ULRtilde=1$, as seen in Fig. \ref{fig:ground_states_sketch} (b). For $\ULRtilde=0$, the Hamiltonian (\ref{eq:H}) without hopping terms scales by the factor two in its on-site repulsion  per site, while the nearest-neighbor and cavity-mediated long-range interaction both scale by the factor four per site. On the other hand for the $\ULRtilde=1$ case, we first rewrite the on-site repulsion term Eq. (\ref{eq:H_U}) as
\begin{equation}
\H_{U}  = \dfrac{U}{2} \sum_i \n_i (\n_i - 1) - U \sum_i \n_{i,\L} \n_{i,\R}
\label{eq:H_U_rewritten}
\end{equation}
and see that the intra-well repulsion Eq. (\ref{eq:H_U_LR}) and last term of Eq. (\ref{eq:H_U_rewritten}) cancel each other out. What remains is the Hamiltonian of a single-well BHM, where on-site repulsion, nearest-neighbor and cavity-mediated long-range interaction scale equally.

\section{Results}
\label{sec:Simulation_Results}
With the QMC-WA, we study the double-well BHM with inter- and intra-well hopping terms to analyze the ground state phase diagrams for various parameter settings. First, we discuss the ground states of the standard double-well BHM (without cavity-mediated long-range and nearest-neighbor interaction). We are interested if the dimerization of hopping terms results in dimerized MI and DW phases with non-integer densities, as expected from the dimerized chain \cite{Mondal2019,Grusdt2013}.\\
Next, the influence of the nearest-neighbor and long-range interaction will be studied. Of special interest for us is hereby the $\rho=1$ lobe. In the one-dimensional single-well BHM, a HI phase occurs at the tip of this lobe, when a nearest-neighbor interaction is present, while a MI phase appears for an included cavity-mediated long-range interaction \cite{Sicks2020}.
\begin{figure*}[t]\centering
\includegraphics[width=2.0\columnwidth]{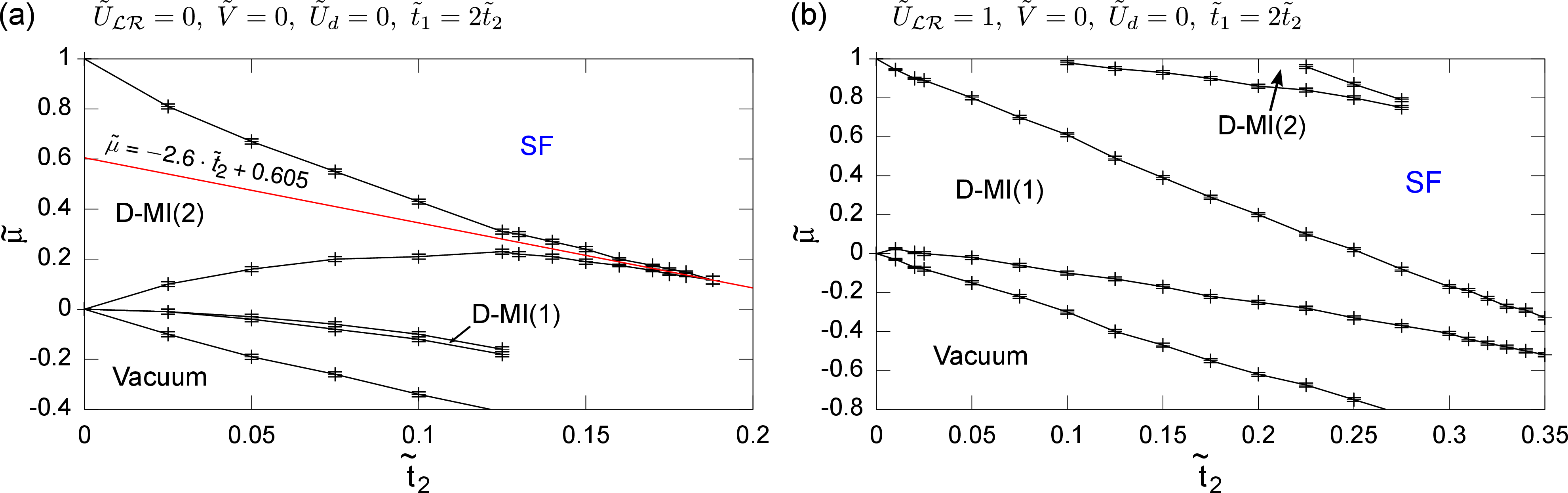}%
\caption{Phase diagram of the standard double-well BHM with $\Vtilde=\Udtilde=0$ and $\ttilde_1 = 2\cdot \ttilde_2$ for (a) $\ULRtilde=0$ and (b) $\ULRtilde=1$. For $\ttilde_2 = 0$, only MI phases are present in both cases, as discussed in Section \ref{sec:Ground_States_without_Hopping_Terms}. (a) When $\ttilde_2>0$, SF phases emerge between Vacuum and D-MI phases, where D-MI phases replace the MI phases. A small D-MI(1) emerges between Vacuum and D-MI(2) phase. The tip of the D-MI(2) lobe is estimated at around $\ttilde_2 \approx 0.22$ with help of the analysis of the order parameters along the constant particle density line $\mutilde = -2.6\cdot\ttilde_2 + 0.605$. (b) When $\ttilde_2>0$, the D-MI phases persist for lower values of $\mutilde$ and stronger hopping strengths until transition into SF phases, compared to the $\ULRtilde=0$ case. This indicates a stabilizing effect of the intra-well hopping for all D-MI(X) phases.}
\label{fig:PD_V_0_Ud_0_t1_2xt2_combined}
\end{figure*}
\begin{figure}[t]\centering
\includegraphics[width=.9\columnwidth]{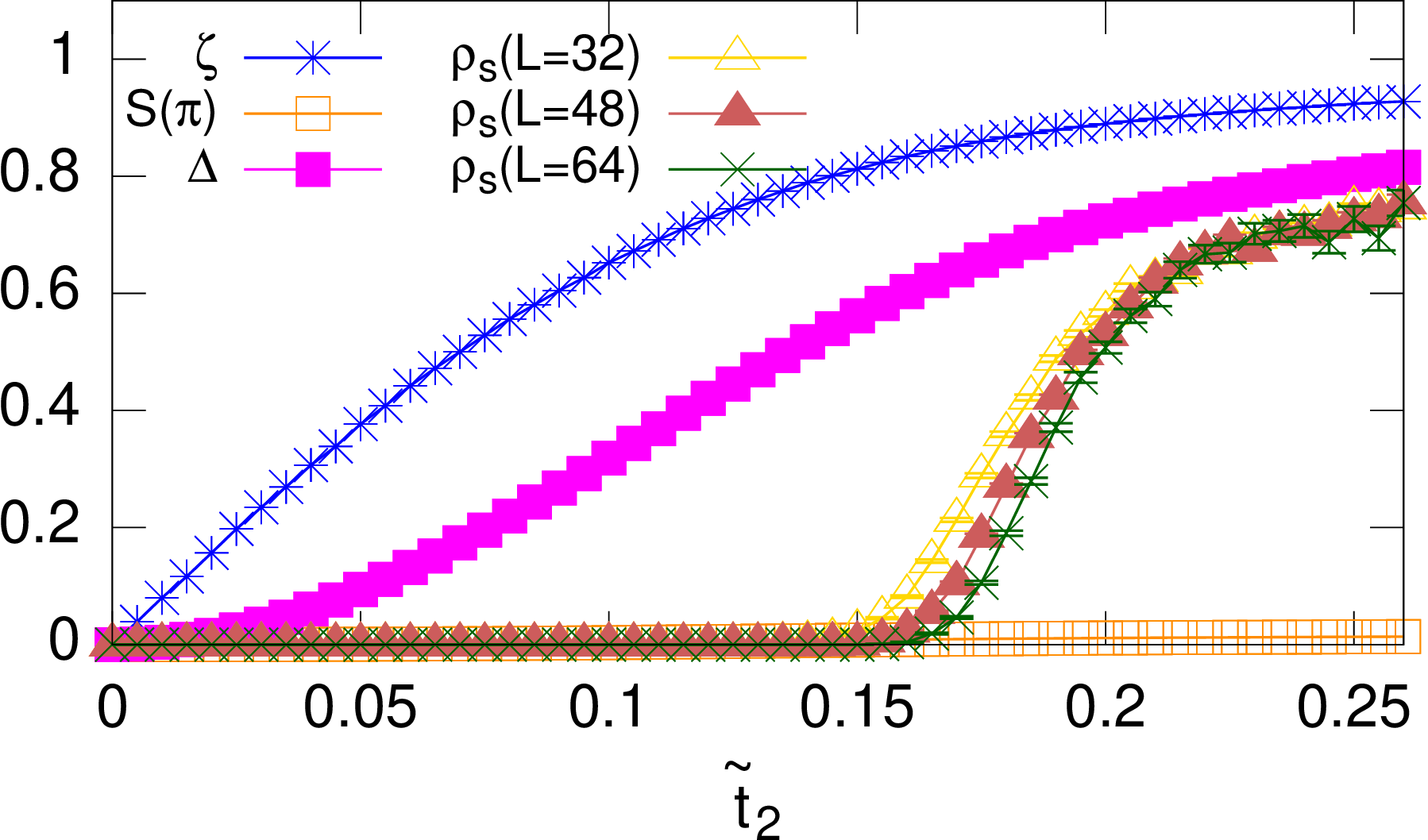}%
\caption{Intra-well fluctuation $\zeta$, structure factor $S(\pi)$, intra-well occupation difference $\Delta$ and superfluid density $\rho_s$ order parameters along the $\mutilde=-2.6\cdot \ttilde_2 + 0.605$ line with constant particle density $\rho=2$ for different chain lengths. While $\rho_s = 0$ the D-MI(2) is present. The tip of the lobe can be approximated by the position, where the superfluid densities of the different chain lengths overlap at ${\ttilde_{2}} \approx 0.22$.}
\label{fig:rho_2_transition}
\end{figure}

\subsection{Standard Double-Well BHM}
\label{subsec:Simulation_Results--Standard Double-Well BHM}

The standard double-well BHM without intra-site repulsion $\ULRtilde=0$ can be interpreted as a single-well dimerized BHM with double chain length (c.f. Fig. \ref{fig:dimerized_chain}). Then, the hopping terms correspond to the dimerized BHM via $t_1=t+\delta$ and $t_2=t-\delta$ with $t$ being the mean hopping strength and $\delta$ the dimerization factor. Therefore, the phase diagram of this parameter setting is expected to be identical to the phase diagram of the dimerized single-well BHM, where dimerized Mott insulator (D-MI) phases emerge, which are a combination of MI and bond-ordered phases \cite{Mondal2019,Kundu2009,Grusdt2013}. The latter appear due to the broken translational symmetry of the dimerized model and is characterized by the alternating strengths of the bond kinetic energy \cite{Nakamura1999}.\\
The phase diagram is depicted in Fig. \ref{fig:PD_V_0_Ud_0_t1_2xt2_combined} (a) and shows the expected behavior. For $\ttilde_2>0$ the MI phases are replaced by D-MI phases, where bosons move between left and right well of the sites. The D-MI(2) phase is hereby identical to the D-MI(1) phase of a single-well dimerized BHM. We calculate the order parameters under the constant density line $\mutilde = -2.6\cdot\ttilde_2 + 0.605$ to analyze the phase transition at the tip of the lobe and show the results in Fig. \ref{fig:rho_2_transition}. For $\rho_s = 0$, the D-MI(2) phase is present, while the transition to the SF phase takes place at around $\ttilde_2 \approx 0.22$, where the superfluid density becomes non-zero and independent of the system size. We compare the phase transition point of our grand-canonical method to the canonical density matrix renormalization group technique used in \cite{Kundu2009}, where the transition point is located around $\ttilde_2 \approx 0.23$ and can confirm that our results match with the single-well dimerized BHM.\\
This includes, that for any dimerization, $\delta \neq 0$, the MI phases are replaced by D-MI phases. Later, this is also shown to be true for the DW and HI phases, which will become D-DW and D-HI phases, respectively. It is because of the higher energy of the intra-well hopping compared to the inter-well hopping. Fluctuations inside the double-wells are more favorable than between neighboring sites which leads to a different kinetic energy contribution between the bond of left and right well of one double-well and the bond between wells of neighboring sites.
\begin{figure*}[t]\centering
\includegraphics[width=2.0\columnwidth]{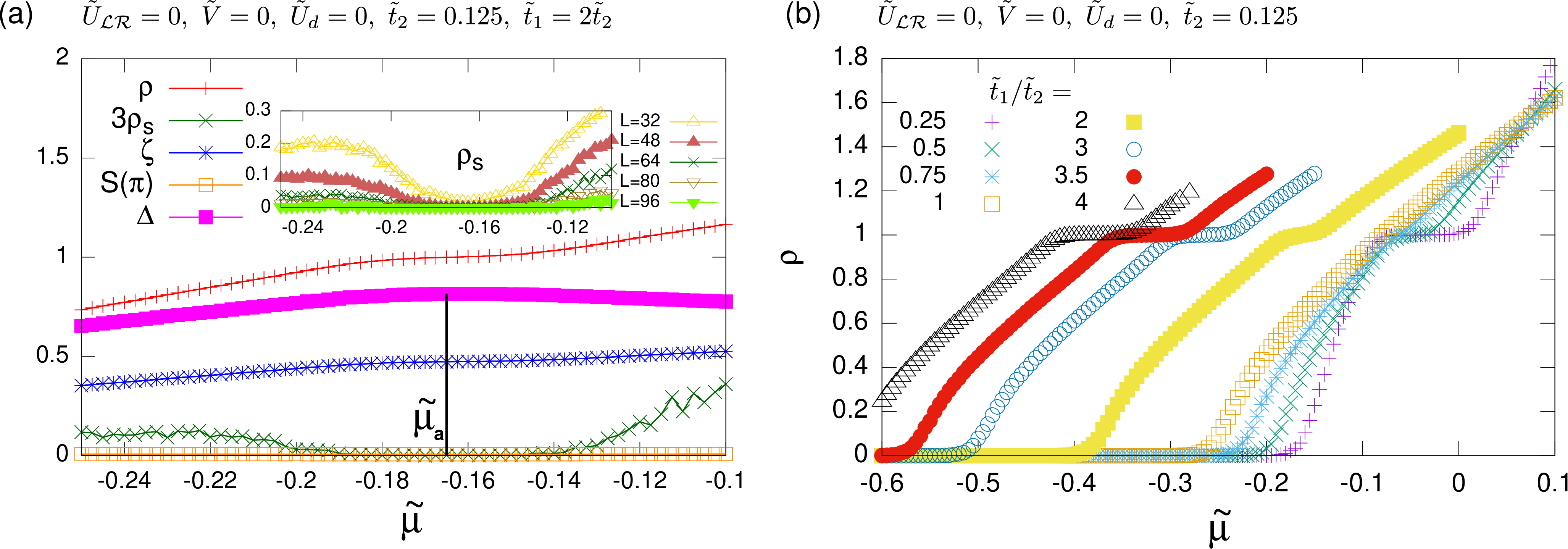}%
\caption{(a) Density $\rho$, superfluid density $\rho_s$, intra-well fluctuation $\zeta$, structure factor $S(\pi)$ and intra-well occupation difference $\Delta$ order parameters for $\ttilde_2 = 0.125$. The position $\mutilde_a \approx -0.165$ is a guide to the eye where $\Delta$ peaks. While $\rho_s$ goes to zero, $\zeta$ remains monotonically increasing. In the inset the behavior of the superfluid density for increasing chain lengths is depicted, where a dip for $\rho=1$ is visible and indicates the presence of a D-MI(1) phase. (b) The particle density $\rho$ over the chemical potential $\mutilde$ for different ratios of $\ttilde_1/\ttilde_2$. A plateau is formed for all dimerized hopping strengths and becomes broader the more the hopping strengths differ in size.}
\label{fig:D-MI_1_phase_combined}
\end{figure*}\\
The existence of a D-MI(1) phase in the standard double-well BHM with $\ULRtilde=0$ is in accordance with the single-well BHM, where no insulating phase exists, when the chain is not dimerized \cite{Kuehner1998}, but a MI-bond-order phase with $\rho=1/2$ emerges, when the chain is dimerized \cite{Mondal2019,Grusdt2013}.\\
In Fig. \ref{fig:D-MI_1_phase_combined}, we investigate this phase in more detail. Beginning from the vacuum state, when the chemical potential is increased, bosons start to occupy the empty chain and can move inside the system by intra-well hopping (between left and right well of each double-well on each site) or inter-well hopping (between left and right wells of double-wells of neighboring sites). When approaching $\rho=1$, nearly all double-wells are occupied by one boson and the well occupation difference $\Delta$ maximizes, as seen at $\mutilde_a$ in Fig. \ref{fig:D-MI_1_phase_combined} (a). As a result, when a boson hops to a double-well of a neighboring site, where another boson is already localized, it hinders the intra-well movement of this boson and cannot hop inside the new double-well itself. This is energetically unfavorable, so the inter-well hopping is suppressed. Only when the chemical potential becomes stronger, this effect will be overcome and the double-well chain populates further with bosons. In the inset of Fig. \ref{fig:D-MI_1_phase_combined} (a) one can see the chain length dependency of the superfluid order parameter, underlining the inter-well hopping decrease around $\rho=1$. Increasing the ratio $\ttilde_1/\ttilde_2$ enhances the aforementioned effect even further and the D-MI(1) phase becomes bigger, as depicted in Fig. \ref{fig:D-MI_1_phase_combined} (b). When $\ttilde_1/\ttilde_2 < 1$, there is a plateau as well, although the intra-well hopping is weaker than the inter-well hopping. This is of no surprise, as a negative bond dimerization $\delta$ only changes the alternating order of the dimerized chain Eq. (\ref{eq:H_t_dim}). The shift of the density with respect to $\mutilde$ is explainable due to the change of the mean hopping strength.\\
~\\
For $\ULRtilde=1$, the energy of a boson being in the same well as another one, is equivalent to a boson being located in the neighboring well on the same site. Hence, the movement of a boson inside a double-well is solely dependent on the intra-well hopping parameter $\ttilde_1$ and not the arrangement of bosons inside the double-well. In Fig. \ref{fig:PD_V_0_Ud_0_t1_2xt2_combined} (b) the phase diagram for the double-well BHM with $\ULRtilde=1$ and $\ttilde_1=2\ttilde_2$ is shown.\\
The resemblance to the single-well BHM \cite{Kuehner1998} is evident, as each double-well in the double-well BHM, for intra-well repulsion strength $\ULRtilde = 1$, behaves in most ways similar to a single-well. The important exception is the intra-well hopping $\ttilde_1$, which introduces more energy due to the movement of bosons inside the double-well and thus leads to a shift of the D-MI phases in the phase diagram to lower $\mutilde$ values and higher hopping values, compared to the single-well BHM phase diagram, where $\ttilde = \frac{3}{2} \ttilde_2$.\\

\subsection{Double-well BHM with Nearest-Neighbor and Long-Range Interactions}
\label{subsec:Simulation_Results--NN and LR Double-Well BHM}

\begin{figure*}[t]\centering
\includegraphics[width=2.0\columnwidth]{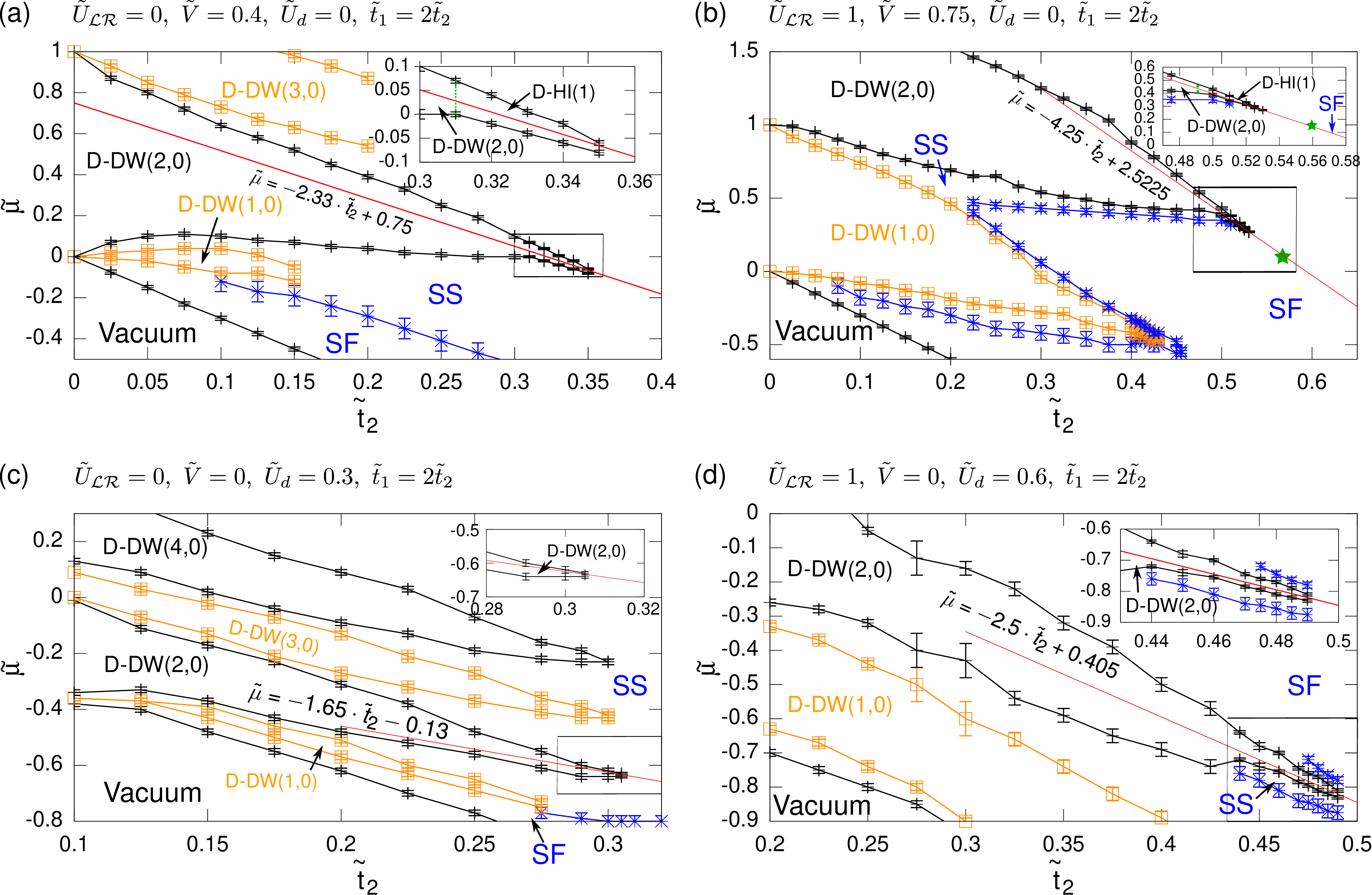}%
\caption{Phase diagrams of the double-well BHM for (a) $\lbrace \ULRtilde=0, \Vtilde=0.4, \Udtilde=0 \rbrace$, (b) $\lbrace 1,0.75,0 \rbrace$, (c) $\lbrace 0,0,0.3 \rbrace$, (d) $\lbrace 1,0,0.6 \rbrace$ and $\ttilde_1 = 2\cdot \ttilde_2$. For $\ttilde_2=0$, only DW phases are present. Phases with non-integer densities are colored orange to better distinguish between the phases and the tips of the D-DW(2,0) lobes, where $\rho=1$, are enhanced in an inset for all phase diagrams. A detailed behavior of the order parameters at the tip for constant densities is carried out along the red lines in Fig. \ref{fig:rho_1_transition_combined}.\\
(a) When $\ttilde_2>0$, D-DW($X_o$,0) phases with odd particle numbers per site $X_o$ appear between D-DW($X_e$,0) phases with even particle numbers. The transition between SF phase and SS phase is depicted via the blue line and expected to engulf the D-DW(1,0) phase completely, so no direct transition from SF to D-DW phase exists. The transitions from SS phases to SF phases for higher values of $\mutilde$ lie at higher hopping values than presented here. At the tip of the D-DW(2,0) phase a D-HI(1) phase is present.\\
(b) When $\ttilde_2>0$, D-DW(X,0) phases emerge and are completely surrounded by SS phases. The transition between SF phase and SS phase is depicted via the blue line. The transitions from SS phases to SF phases for higher values of $\mutilde$ are not presented here. A D-HI(1) phase appears at the tip of the D-DW(2,0) lobe and a transition from D-HI(1) to SF can be determined at the position of the green star.\\
(c) D-DW($X_e$,0) phases, where $X_e$ is even, appear as soon as hopping is included. In between, D-DW($X_o$,0) phases, with $X_o$ being odd, emerge due to the stabilizing effect of the intra-well hopping. The transition from SF to SS phase is depicted via the blue line and is expected to engulf the D-DW(1,0) phase. Transitions from SS to SF phases for higher values of $\mutilde$ are not presented here. \\
(d) D-DW(X,0) phases appear, when the hopping terms are included. The D-DW phases are surrounded by a narrow SS phase, but only the transition at the tip is depicted via the blue line. At the tip of the D-DW(2,0) phase a transition to a D-MI(1) phase can be seen.
}
\label{fig:PD_t1_2xt2_combined}
\end{figure*}

We study the nearest-neighbor and cavity-mediated long-range interaction for different parameter settings and show the phase diagrams in Fig. \ref{fig:PD_t1_2xt2_combined}. We have chosen the nearest-neighbor and long-range interaction strengths to be in the regime, where for $\ttilde_2=0$ only DW phases are present (see Fig. \ref{fig:ground_states_sketch} (b)). We compare our results with the standard double-well BHM and the (dimerized) single-well BHM. For the nearest-neighbor interaction, at the tip of the DW(2,0) lobe a HI phase was found in the undimerized case \cite{Batrouni2013} and a D-HI in the dimerized BHM \cite{Sugimoto2019}. For the single-well BHM with cavity-mediated long-range interaction a MI phase is located at the tip of the DW(2,0) lobe \cite{Sicks2020}.
\\
Regarding all phase diagrams presented in Fig. \ref{fig:PD_t1_2xt2_combined}, when the hopping is greater than zero, MI, DW and HI phases will become dimerized phases, meaning that bosons are localized in the double-well on one site and fluctuate between left and right well of this double-well.
\begin{figure*}[t]\centering
\includegraphics[width=2.0\columnwidth]{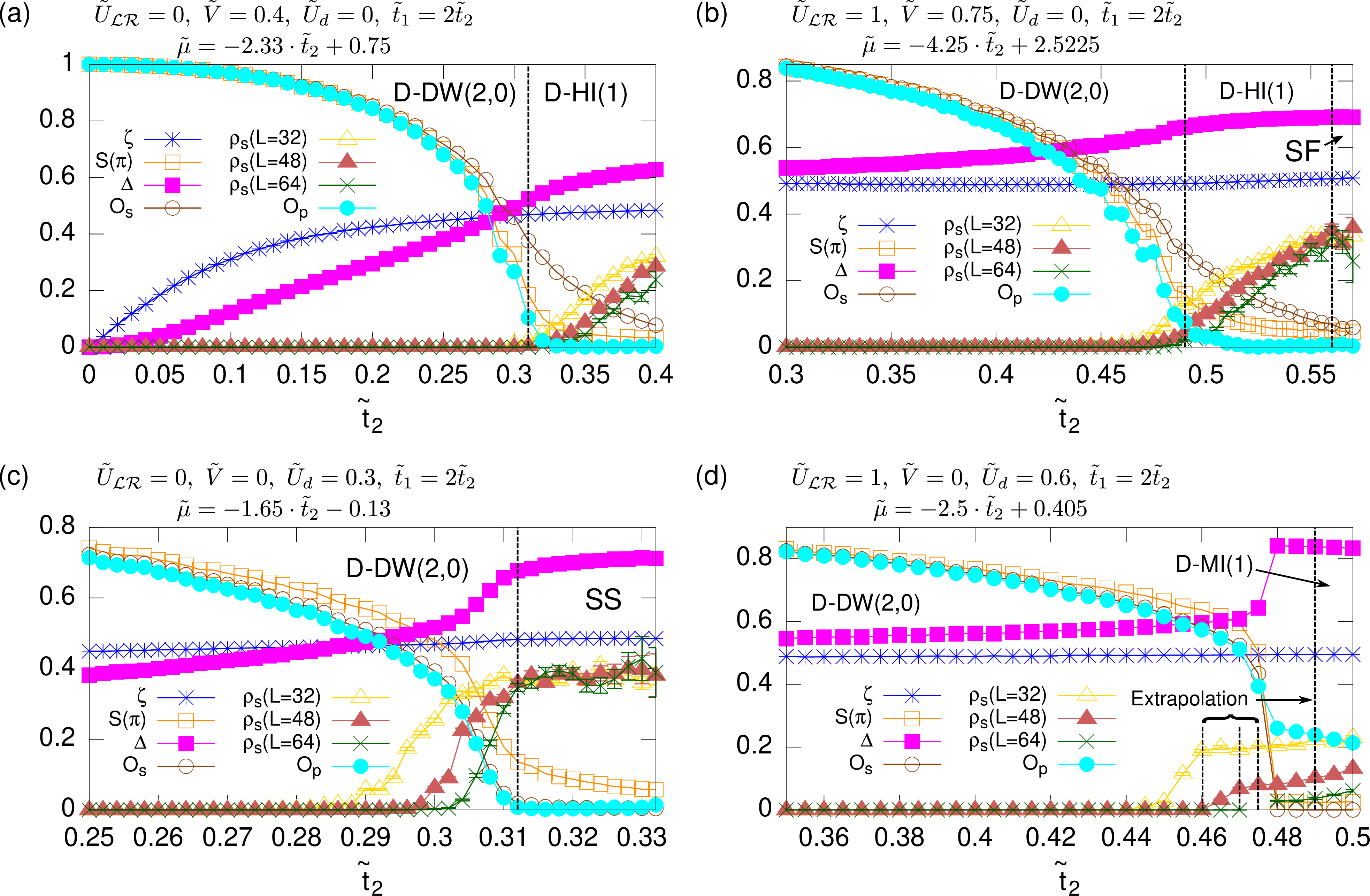}%
\caption{Intra-well fluctuation $\zeta$, structure factor $S(\pi)$, intra-well occupation difference $\Delta$, superfluid density $\rho_s$, string $\O_s$ and parity $\O_p$ order parameters along the constant density $\rho=1$ lines (a) $\mutilde=-2.33\cdot\ttilde_2 + 0.75$ for $\lbrace \ULRtilde=0, \Vtilde=0.4, \Udtilde=0 \rbrace$, (b) $\mutilde=-4.25\cdot\ttilde_2 + 2.5225$ for  $\lbrace 1,0.75,0 \rbrace$, (c) $\mutilde=-1.65\cdot\ttilde_2 - 0.13$ for $\lbrace 0,0,0.3 \rbrace$, (d) $\mutilde=-2.5\cdot\ttilde_2 + 0.405$ for  $\lbrace 1,0,0.6 \rbrace$ and $\ttilde_1 = 2\cdot \ttilde_2$. In the D-DW(2,0) phase, only the superfluid density is zero, while all other order parameters are non-zero.\\
(a) At the transition point to the D-HI(1) phase at around $\ttilde_2 \approx 0.31$, $S(\pi)$ and $\O_p$ vanish, while $\O_s$ keeps a finite value. The superfluid density $\rho_s$ attains non-zero values, but approaches zero for larger chain lengths.\\
(b) At the transition point to the D-HI(1) phase at around $\ttilde_2 \approx 0.49$, $S(\pi)$ and $\O_p$ vanish, while $\O_s$ keeps a finite value. The superfluid density $\rho_s$ attains non-zero values, but approaches zero for larger chain lengths. The transition to the superfluid phase can be determined at $\ttilde=0.56$, when $\rho_s$ becomes size-independent and $\O_s$ vanishes.\\
(c) At the transition point $\ttilde_2 \approx 0.312$ the SS phase appears, as $\O_s$ and $\O_p$ vanish, while $S(\pi)$ remains present. No dimerized HI or MI phase can be identified between the D-DW(2,0) and SS phases.\\
(d) At the transition point $\ttilde_2 \approx 0.49$ a D-MI(1) phase appears, as $\O_s$ and $S(\pi)$ vanish, while $\O_s$ remains present and $\rho_s$ approaches zero for larger chain lengths. Because the point, where the superfluid density becomes non-zero, is highly dependent on the system size, we extrapolated it for the different sizes to obtain a better approximation for the transition point to the D-MI(1) phase.
}
\label{fig:rho_1_transition_combined}
\end{figure*}\\
For $\ULRtilde = 0$, D-DW($X_o$,0) phases emerge in between integer density phases for nearest-neighbor Fig. \ref{fig:PD_t1_2xt2_combined} (a) and long-range Fig. \ref{fig:PD_t1_2xt2_combined} (c) interactions. The reason for this is analogue to the D-MI(1) phase in the standard double-well BHM in Fig. \ref{fig:PD_V_0_Ud_0_t1_2xt2_combined} (a). Additionally, the D-DW phases are carried out for bigger hopping values. When $\Vtilde > 0$, the SF and SS phases are shifted to higher energy values, while the D-DW phases are unaffected by the nearest-neighbor term and thus persist longer as in the standard double-well BHM. For the long-range interaction, when $\Udtilde>0$, the argumentation reverses. The energy of the D-DW phases are lowered, while the SF and SS phases are unaffected. Hence, not only do the D-DW phases persist for bigger hopping terms, but also for smaller values of $\mutilde$.\\
For $\ULRtilde = 1$, the argumentation, that D-DW phases are present at higher hopping strengths and shift to lower $\mutilde$ values for the long-range interaction compared to the $\ULRtilde=0$ case, remains the same for the phase diagrams with nearest-neighbor Fig. \ref{fig:PD_t1_2xt2_combined} (b) and long-range Fig. \ref{fig:PD_t1_2xt2_combined} (d) interaction. Yet in contrast to the $\ULRtilde=0$ case, the D-DW($X_o$,0) phases become broader when they approach $\ttilde_2=0$.\\
For the extended double-well BHM Fig. \ref{fig:PD_t1_2xt2_combined} (a) and (b), at the tip of the D-DW(2,0) phase, a D-HI(1) phase can be identified, while for the long-range interaction Fig. \ref{fig:PD_t1_2xt2_combined} (c) and (d) only at the tip of the D-DW(2,0) lobe for the $\ULRtilde=1,~\Udtilde=0.6$ diagram, a D-MI(1) phase was extrapolated.\\
The order parameters along the constant density lines for the D-DW(2,0) phase of the phase diagrams in Fig. \ref{fig:PD_t1_2xt2_combined} are given in Fig. \ref{fig:rho_1_transition_combined}. The labels (a)-(d) correspond in both figures. On the left hand side of each picture is the D-DW(2,0) phase, where only the superfluid density $\rho_s$ is zero, while all other order parameters obtain non-zero values.\\
For the extended double-well BHM in Fig. \ref{fig:rho_1_transition_combined} (a) and (b) the D-DW(2,0) phase transits into a D-HI(1) phase, where structure factor $S(\pi)$ and parity order parameter $\O_p$ drop to zero, while the string order parameter $\O_s$ remains non-zero. The superfluid density becomes non-zero, but is size-dependent and becomes zero in the limit $L\rightarrow \infty$. Furthermore, in Fig. \ref{fig:rho_1_transition_combined} (b) the transition to the SF can also be seen, where $\rho_s$ is size-independent and non-zero, while $\O_s$ approaches zero. Comparing these results for the D-HI phase with the results for the single-well extended BHM \cite{Batrouni2013,Sicks2020} shows, that the topological HI phase persists in the double-well BHM in a dimerized way, where the intra-well hopping does not break the long-range order of the HI phase.\\
Since the nearest-neighbor interaction of our double-well BHM is identical to the dimerized chain where the nearest-neighbor interaction acts on neighboring pairs of sites, we showed that the D-HI phase persists also for the dimerized chain with interactions up to the third neighboring site. This is an expansion of the results of Sugimoto \textit{et al.} \cite{Sugimoto2019}, where a D-HI phase was found in the dimerized chain with nearest-neighbor interaction.\\
For the cavity-mediated long-range interaction with $\ULRtilde=0,~\Udtilde=0.3$, shown in Fig. \ref{fig:rho_1_transition_combined} (c), the D-DW(2,0) directly transits into the SS phase, where $\rho_s$ and $S(\pi)$ are non-zero and $\O_s$ and $\O_p$ zero.
To understand why there exists no D-MI(1) phase at the tip of the D-DW(2,0) lobe, contrary to the single-well BHM, where a MI phase is present at the tip of the DW(2,0) lobe, we reiterate why the MI(1) phase exists in the single-well BHM in the first place.\\
The MI(1) phase appears at the tip of the DW(2,0) lobe in the single-well BHM because the inter-site particle fluctuation per site scales in the MI phase by the power of $U_d/L$, as particles are evenly distributed between even and odd sites. On the other hand, inter-site particle fluctuation per site for the DW(X,0) phases scale by the factor $X^2U_d/4$, making it independent of system size. This argumentation holds true also for the double-well BHM, but is expanded with the effect of the intra-well hopping, where the bosons in the D-DW(2,0) phase are able to fluctuate inside every second double-well. This overcomes the effects of the cavity-mediated long-range interaction on the inter-well fluctuations for the D-DW(2,0) phase and results in the D-DW(2,0) to be present until the hopping strengths are strong enough that a direct transition into the SS phase occurs.\\
For the $\ULRtilde=1,~\Udtilde=0.6$ case, shown in Fig. \ref{fig:rho_1_transition_combined} (d), a transition to the D-MI(1) phase can be seen, where $\O_p$ persists, while $\rho_s$ goes to zero for larger system sizes. Due to the strong variance of the starting point, where the superfluid density becomes non-zero according to system size, an extrapolation was carried out to determine the position of the transition to the D-MI(1) phase. The occurrence of a D-MI(1) at the tip of the D-DW(2,0) lobe in the double-well BHM with long-range interaction $\Udtilde=0.6$ matches with the finding of a MI(1) phase at the tip of the DW(2,0) lobe in the single-well BHM with cavity-mediated long-range interaction \cite{Sicks2020}. This is reasonable, as we have showed in Section \ref{sec:Ground_States_without_Hopping_Terms} for the double-well BHM without hopping, that the double-well BHM with $\ULRtilde=1$ is identical to a single-well BHM. Our results confirm, that a dimerization of the hopping keeps the structure of a MI(1) phase at the tip of the DW(2,0), but dimerizes both phases to D-MI(1) and D-DW(2,0) respectively.

\section{Conclusion}
\label{sec:Conclusion}

The one-dimensional double-well BHM with nearest-neighbor and cavity-mediated long-range interaction includes a variety of well-established models, like the single-well BHM and the dimerized BHM. When dimerization is present, dimerized Mott insulator (D-MI), dimerized Density Wave (D-DW) and dimerized Haldane insulator (D-HI) phases exist, characterized by a combination of a bond-ordered phase \cite{Nakamura1999} and a MI, DW and HI phase, respectively. It turns out that dimerized hopping stabilizes the D-DW phases with non-integer boson densities, in agreement with results from dimerized chains \cite{Mondal2019,Grusdt2013}. When the intra-well repulsion is as strong as the on-site repulsion per well, each double-well can be treated as a single-well. Hence in this case, the ground state phase diagram is identical to the single-well BHM \cite{Kuehner1998}, with the exception, that the intra-well hopping leads to dimerized phases and a shift of the D-MI phases to lower chemical potentials and higher hopping values.\\
Interactions between particles in neighboring double-well potentials imply interactions up to the third neighbor in the corresponding dimerized BHM. For those a dimerized Haldane insulator, D-HI, phase at the tip of the D-DW lobe with particle density one exists, which was previously reported for a dimerized chain with solely nearest-neighbor interaction \cite{Sugimoto2019}. For non-integer densities D-DW phases exist, as was recently shown for the dimerized BH chain with nearest-neighbor interactions \cite{Hayashi2022}.\\
In the presence of cavity-mediated long-range interactions a D-MI phase appears at the tip of the D-DW lobe with density one when the intra-well repulsion is as strong as the on-site repulsion, which is in agreement with results for the single-well chain BHM \cite{Sicks2020}. This D-MI phase at the tip of the D-DW phase disappears when the the intra-well repulsion vanishes.  This is due to the intra-well hopping, which distinguishes the D-DW phase in the double-well BHM and the DW phase of the single-well BHM. Moreover, D-DW phases exist for non-integer densities in the double-well BHM with cavity-mediated long-range interaction, which is reminiscent of the double-well BHM with nearest-neighbor interaction and underpins the equivalence of nearest-neighbor and cavity-mediated long-range interaction on a mean-field level \cite{Dogra2016}.

%
%% For one-column wide figures use
%\begin{figure}
%% Use the relevant command for your figure-insertion program
%% to insert the figure file.
%% For example, with the option graphics use
%\resizebox{0.75\textwidth}{!}{%
%  \includegraphics{leer.eps}
%}
%% If not, use
%%\vspace{5cm}       % Give the correct figure height in cm
%\caption{Please write your figure caption here}
%\label{fig:1}       % Give a unique label
%\end{figure}
%%
%% For two-column wide figures use
%\begin{figure*}
%% Use the relevant command for your figure-insertion program
%% to insert the figure file. See example above.
%% If not, use
%\vspace*{5cm}       % Give the correct figure height in cm
%\caption{Please write your figure caption here}
%\label{fig:2}       % Give a unique label
%\end{figure*}
%%
%% For tables use
%\begin{table}
%\caption{Please write your table caption here}
%\label{tab:1}       % Give a unique label
%% For LaTeX tables use
%\begin{tabular}{lll}
%\hline\noalign{\smallskip}
%first & second & third  \\
%\noalign{\smallskip}\hline\noalign{\smallskip}
%number & number & number \\
%number & number & number \\
%\noalign{\smallskip}\hline
%\end{tabular}
%% Or use
%\vspace*{5cm}  % with the correct table height
%\end{table}
%
% The section below may be edited at your convenience to acknowledge 
% each author's contribution to the manuscript.
% You may remove it if you are a single author.
%
%\section{Authors contributions}
%All the authors were involved in the preparation of the manuscript.
%All the authors have read and approved the final manuscript.
%

\section*{Appendix A: Quantum Monte-Carlo Worm-Algorithm}
\label{app:A}
In this appendix we discuss the quantum Monte Carlo worm-algorithm we used to obtain the phase diagram of Hamiltonian (\ref{eq:H}) in more detail. We split the Hamiltonian in an on-diagonal part $\Hon$ and off-diagonal part $\Hoff$ with regard to the Fock basis representation of the 1D chain $\ket{\textbf{n}_i} = \ket{n_1 \dots n_L}_i$. Hence, $\Hon \ket{\textbf{n}_i}$ gives the on-diagonal energy value $\epsilon_i$. With the inverse temperature $\beta$ and the Dyson series, we can write the partition function as
\begin{align*}
\Z(\C) = &\sum_{m=0}^{\infty} \sum_{\textbf{n}_1 \dots \textbf{n}_m} e^{-\beta \epsilon_1} \int_0^\beta d\tau_m \cdots \int_{0}^{\tau_{2}} d\tau_1 \\ \nonumber
&\times\l( e^{\tau_m \e_1} \Hoff^{\textbf{n}_1 \textbf{n}_m} e^{-\tau_m \e_m} \r) \cdots \l( e^{\tau_1 \e_2} \Hoff^{\textbf{n}_2 \textbf{n}_1} e^{-\tau_1 \e_1} \r) ,
\label{eq:Z}
\tag{A1}
\end{align*}
with $\Hoff^{\textbf{n}_i \textbf{n}_j}= \braket{\textbf{n}_i | \Hoff | \textbf{n}_j}$.\\
The partition function is the sum over all possible configurations, where $m$ denotes the number of vertices in the system, as it counts the number of $\Hoff$ terms. The imaginary time dimension ranges from $[0,\beta]$ and is segmented by the vertices in different lengths $\tau_m$. Additionally, the Fock states at the beginning and the end must be the same $\ket{\textbf{n}_0} = \ket{\textbf{n}_m}$.
\begin{figure}[t]\centering
\includegraphics[width=\columnwidth]{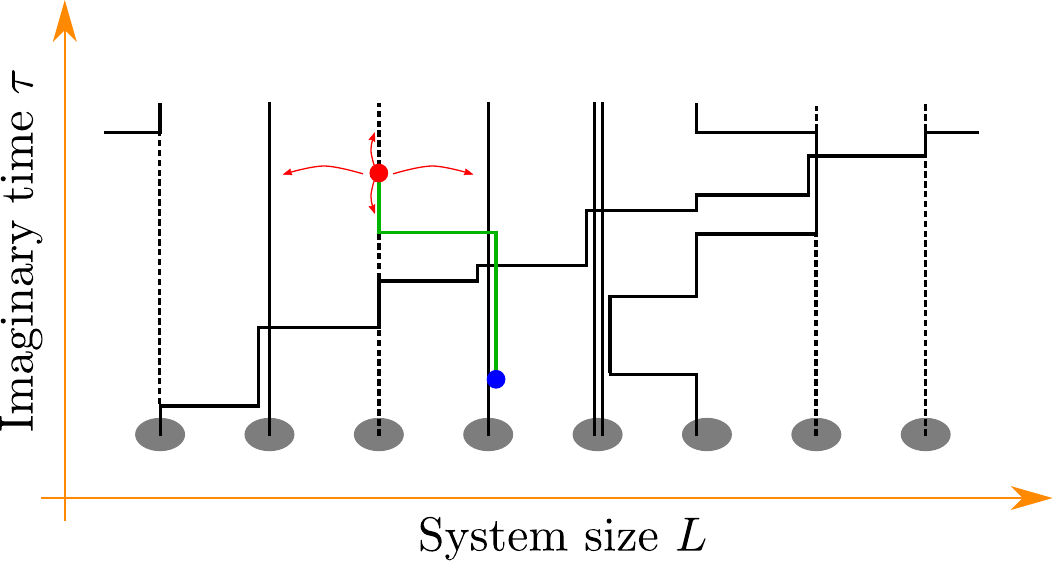}%
\caption{Schematic movement of a worm inside the 1D chain. Solid black lines represent closed boson lines. Dashed black lines show vacancies. The open boson line is coloured green and has the worm head (red) and tail (blue) as ends. The ends can move through the system as implied by the red arrows for the worm head. When head and tail collide, the worm closes, leaving new closed boson lines behind. Note, that when at one site more than one boson is present, any of these bosons may perform a hopping, as they are all indistinguishable. The depiction in the figure was chosen to easily tell the occupation number per site. Also the boson configuration at the start $\tau=0$ must be the same as at the end $\tau=\beta$.}
\label{fig:1d_bose_hubbard}
\end{figure}\\
Now, the worm is inserted by including a $\b^{\dagger~\textbf{n}'_c \textbf{n}_c}_i (\tau_c)~\b^{\textbf{n}'_a \textbf{n}_a}_i (\tau_a)$ or $\b^{\textbf{n}'_a \textbf{n}_a}_i (\tau_a)~\b^{\dagger~\textbf{n}'_c \textbf{n}_c}_i (\tau_c)$ pair at an arbitrary site $i$ and imaginary time position $\tau_c=\tau_a$. In the first case, between the operators, a boson is removed, while in the second case a boson is created. Obviously, the worm is only inserted, if a boson can be removed, or the maximal amount of bosons on one site is not surpassed, when there is such a limit defined.\\
So, the extended configuration for the partition function kernel reads
\begin{align*}
\C^{\text{ext}} &=\l( e^{\tau_m \e_1} \Hoff^{\textbf{n}_1 \textbf{n}_m} e^{-\tau_m \e_m} \r) \cdots \l( e^{\tau_c \e'_c} \b^{\dagger~\textbf{n}'_c \textbf{n}_c}_i e^{-\tau_c \e_c} \r) \cdots\\ \nonumber
&\times \l( e^{\tau_a \e'_a} \b^{\textbf{n}'_a \textbf{n}_a}_i e^{-\tau_a \e_a} \r) \cdots \l( e^{\tau_1 \e_2} \Hoff^{\textbf{n}_2 \textbf{n}_1} e^{-\tau_1 \e_1} \r) 
\label{eq:Z}
\tag{A2}
\end{align*}
and vice versa for the other pairing. W.l.o.g. we define the creator operator as head and the annihilator operator as tail. They can move through the configuration space by advancing forwards and backwards in imaginary time, or by hopping accordingly to the off-diagonal term $\Hoff$. When they come across an already existing vertex, where they cannot pass through (because the commutator does not vanish), one of three scenarios happens. Either the vertex gets deleted, the vertex is relinked to another site according to $\Hoff$, or nothing happens and the worm end moves in the other direction. The worm movement ends when head and tail collide. Fig. \ref{fig:1d_bose_hubbard} sketches the 1D quantum chain, expanded with the imaginary time to a classical 2D model and the insertion of a worm.\\
With the QMC-WA it is possible to obtain grand-canonical and canonical observables. During worm movements the boson number is variable, allowing directly to obtain the Green's function for example. When the worm is deleted, the particle number is constant and canonical observables can be calculated. The usual approach is via importance sampling
\begin{equation}
\langle \O \rangle = \dfrac{1}{\Z} \sum_\C \O(\C) \Z(\C).
\tag{A3}
\end{equation}

\bibliography{bib_double-well_BHM}

%apsrev4-2.bst 2019-01-14 (MD) hand-edited version of apsrev4-1.bst
%Control: key (0)
%Control: author (72) initials jnrlst
%Control: editor formatted (1) identically to author
%Control: production of article title (-1) disabled
%Control: page (0) single
%Control: year (1) truncated
%Control: production of eprint (0) enabled
\begin{thebibliography}{64}%
\makeatletter
\providecommand \@ifxundefined [1]{%
 \@ifx{#1\undefined}
}%
\providecommand \@ifnum [1]{%
 \ifnum #1\expandafter \@firstoftwo
 \else \expandafter \@secondoftwo
 \fi
}%
\providecommand \@ifx [1]{%
 \ifx #1\expandafter \@firstoftwo
 \else \expandafter \@secondoftwo
 \fi
}%
\providecommand \natexlab [1]{#1}%
\providecommand \enquote  [1]{``#1''}%
\providecommand \bibnamefont  [1]{#1}%
\providecommand \bibfnamefont [1]{#1}%
\providecommand \citenamefont [1]{#1}%
\providecommand \href@noop [0]{\@secondoftwo}%
\providecommand \href [0]{\begingroup \@sanitize@url \@href}%
\providecommand \@href[1]{\@@startlink{#1}\@@href}%
\providecommand \@@href[1]{\endgroup#1\@@endlink}%
\providecommand \@sanitize@url [0]{\catcode `\\12\catcode `\$12\catcode
  `\&12\catcode `\#12\catcode `\^12\catcode `\_12\catcode `\%12\relax}%
\providecommand \@@startlink[1]{}%
\providecommand \@@endlink[0]{}%
\providecommand \url  [0]{\begingroup\@sanitize@url \@url }%
\providecommand \@url [1]{\endgroup\@href {#1}{\urlprefix }}%
\providecommand \urlprefix  [0]{URL }%
\providecommand \Eprint [0]{\href }%
\providecommand \doibase [0]{https://doi.org/}%
\providecommand \selectlanguage [0]{\@gobble}%
\providecommand \bibinfo  [0]{\@secondoftwo}%
\providecommand \bibfield  [0]{\@secondoftwo}%
\providecommand \translation [1]{[#1]}%
\providecommand \BibitemOpen [0]{}%
\providecommand \bibitemStop [0]{}%
\providecommand \bibitemNoStop [0]{.\EOS\space}%
\providecommand \EOS [0]{\spacefactor3000\relax}%
\providecommand \BibitemShut  [1]{\csname bibitem#1\endcsname}%
\let\auto@bib@innerbib\@empty
%</preamble>
\bibitem [{\citenamefont {Fisher}\ \emph {et~al.}(1989)\citenamefont {Fisher},
  \citenamefont {Weichman}, \citenamefont {Grinstein},\ and\ \citenamefont
  {Fisher}}]{Fisher1989}%
  \BibitemOpen
  \bibfield  {author} {\bibinfo {author} {\bibfnamefont {M.~P.~A.}\
  \bibnamefont {Fisher}}, \bibinfo {author} {\bibfnamefont {P.~B.}\
  \bibnamefont {Weichman}}, \bibinfo {author} {\bibfnamefont {G.}~\bibnamefont
  {Grinstein}},\ and\ \bibinfo {author} {\bibfnamefont {D.~S.}\ \bibnamefont
  {Fisher}},\ }\href {https://doi.org/10.1103/PhysRevB.40.546} {\bibfield
  {journal} {\bibinfo  {journal} {Phys. Rev. B}\ }\textbf {\bibinfo {volume}
  {40}},\ \bibinfo {pages} {546} (\bibinfo {year} {1989})}\BibitemShut
  {NoStop}%
\bibitem [{\citenamefont {Jaksch}\ \emph {et~al.}(1998)\citenamefont {Jaksch},
  \citenamefont {Bruder}, \citenamefont {Cirac}, \citenamefont {Gardiner},\
  and\ \citenamefont {Zoller}}]{Jaksch1998}%
  \BibitemOpen
  \bibfield  {author} {\bibinfo {author} {\bibfnamefont {D.}~\bibnamefont
  {Jaksch}}, \bibinfo {author} {\bibfnamefont {C.}~\bibnamefont {Bruder}},
  \bibinfo {author} {\bibfnamefont {J.~I.}\ \bibnamefont {Cirac}}, \bibinfo
  {author} {\bibfnamefont {C.~W.}\ \bibnamefont {Gardiner}},\ and\ \bibinfo
  {author} {\bibfnamefont {P.}~\bibnamefont {Zoller}},\ }\href
  {https://doi.org/10.1103/PhysRevLett.81.3108} {\bibfield  {journal} {\bibinfo
   {journal} {Phys. Rev. Lett.}\ }\textbf {\bibinfo {volume} {81}},\ \bibinfo
  {pages} {3108} (\bibinfo {year} {1998})}\BibitemShut {NoStop}%
\bibitem [{\citenamefont {Greiner}\ \emph {et~al.}(2002)\citenamefont
  {Greiner}, \citenamefont {Mandel}, \citenamefont {Esslinger}, \citenamefont
  {H{\"a}nsch},\ and\ \citenamefont {Bloch}}]{Greiner2002}%
  \BibitemOpen
  \bibfield  {author} {\bibinfo {author} {\bibfnamefont {M.}~\bibnamefont
  {Greiner}}, \bibinfo {author} {\bibfnamefont {O.}~\bibnamefont {Mandel}},
  \bibinfo {author} {\bibfnamefont {T.}~\bibnamefont {Esslinger}}, \bibinfo
  {author} {\bibfnamefont {T.~W.}\ \bibnamefont {H{\"a}nsch}},\ and\ \bibinfo
  {author} {\bibfnamefont {I.}~\bibnamefont {Bloch}},\ }\href
  {https://doi.org/10.1038/415039a} {\bibfield  {journal} {\bibinfo  {journal}
  {Nature}\ }\textbf {\bibinfo {volume} {415}},\ \bibinfo {pages} {39}
  (\bibinfo {year} {2002})}\BibitemShut {NoStop}%
\bibitem [{\citenamefont {Landig}\ \emph {et~al.}(2016)\citenamefont {Landig},
  \citenamefont {Hruby}, \citenamefont {Dogra}, \citenamefont {Landini},
  \citenamefont {Mottl}, \citenamefont {Donner},\ and\ \citenamefont
  {Esslinger}}]{Landig2016}%
  \BibitemOpen
  \bibfield  {author} {\bibinfo {author} {\bibfnamefont {R.}~\bibnamefont
  {Landig}}, \bibinfo {author} {\bibfnamefont {L.}~\bibnamefont {Hruby}},
  \bibinfo {author} {\bibfnamefont {N.}~\bibnamefont {Dogra}}, \bibinfo
  {author} {\bibfnamefont {M.}~\bibnamefont {Landini}}, \bibinfo {author}
  {\bibfnamefont {R.}~\bibnamefont {Mottl}}, \bibinfo {author} {\bibfnamefont
  {T.}~\bibnamefont {Donner}},\ and\ \bibinfo {author} {\bibfnamefont
  {T.}~\bibnamefont {Esslinger}},\ }\href
  {http://dx.doi.org/10.1038/nature17409} {\bibfield  {journal} {\bibinfo
  {journal} {Nature}\ }\textbf {\bibinfo {volume} {532}},\ \bibinfo {pages}
  {476 EP } (\bibinfo {year} {2016})}\BibitemShut {NoStop}%
\bibitem [{\citenamefont {Anderlini}\ \emph {et~al.}(2006)\citenamefont
  {Anderlini}, \citenamefont {Sebby-Strabley}, \citenamefont {Kruse},
  \citenamefont {Porto},\ and\ \citenamefont {Phillips}}]{Anderlini2006}%
  \BibitemOpen
  \bibfield  {author} {\bibinfo {author} {\bibfnamefont {M.}~\bibnamefont
  {Anderlini}}, \bibinfo {author} {\bibfnamefont {J.}~\bibnamefont
  {Sebby-Strabley}}, \bibinfo {author} {\bibfnamefont {J.}~\bibnamefont
  {Kruse}}, \bibinfo {author} {\bibfnamefont {J.~V.}\ \bibnamefont {Porto}},\
  and\ \bibinfo {author} {\bibfnamefont {W.~D.}\ \bibnamefont {Phillips}},\
  }\href {https://doi.org/10.1088/0953-4075/39/10/s19} {\bibfield  {journal}
  {\bibinfo  {journal} {Journal of Physics B: Atomic, Molecular and Optical
  Physics}\ }\textbf {\bibinfo {volume} {39}},\ \bibinfo {pages} {S199}
  (\bibinfo {year} {2006})}\BibitemShut {NoStop}%
\bibitem [{\citenamefont {Anderlini}\ \emph {et~al.}(2007)\citenamefont
  {Anderlini}, \citenamefont {Lee}, \citenamefont {Brown}, \citenamefont
  {Sebby-Strabley}, \citenamefont {Phillips},\ and\ \citenamefont
  {Porto}}]{Anderlini2007}%
  \BibitemOpen
  \bibfield  {author} {\bibinfo {author} {\bibfnamefont {M.}~\bibnamefont
  {Anderlini}}, \bibinfo {author} {\bibfnamefont {P.~J.}\ \bibnamefont {Lee}},
  \bibinfo {author} {\bibfnamefont {B.~L.}\ \bibnamefont {Brown}}, \bibinfo
  {author} {\bibfnamefont {J.}~\bibnamefont {Sebby-Strabley}}, \bibinfo
  {author} {\bibfnamefont {W.~D.}\ \bibnamefont {Phillips}},\ and\ \bibinfo
  {author} {\bibfnamefont {J.~V.}\ \bibnamefont {Porto}},\ }\href
  {https://doi.org/10.1038/nature06011} {\bibfield  {journal} {\bibinfo
  {journal} {Nature}\ }\textbf {\bibinfo {volume} {448}},\ \bibinfo {pages}
  {452} (\bibinfo {year} {2007})}\BibitemShut {NoStop}%
\bibitem [{\citenamefont {Atala}\ \emph {et~al.}(2013)\citenamefont {Atala},
  \citenamefont {Aidelsburger}, \citenamefont {Barreiro}, \citenamefont
  {Abanin}, \citenamefont {Kitagawa}, \citenamefont {Demler},\ and\
  \citenamefont {Bloch}}]{Atala2013}%
  \BibitemOpen
  \bibfield  {author} {\bibinfo {author} {\bibfnamefont {M.}~\bibnamefont
  {Atala}}, \bibinfo {author} {\bibfnamefont {M.}~\bibnamefont {Aidelsburger}},
  \bibinfo {author} {\bibfnamefont {J.~T.}\ \bibnamefont {Barreiro}}, \bibinfo
  {author} {\bibfnamefont {D.}~\bibnamefont {Abanin}}, \bibinfo {author}
  {\bibfnamefont {T.}~\bibnamefont {Kitagawa}}, \bibinfo {author}
  {\bibfnamefont {E.}~\bibnamefont {Demler}},\ and\ \bibinfo {author}
  {\bibfnamefont {I.}~\bibnamefont {Bloch}},\ }\href
  {https://doi.org/10.1038/nphys2790} {\bibfield  {journal} {\bibinfo
  {journal} {Nature Physics}\ }\textbf {\bibinfo {volume} {9}},\ \bibinfo
  {pages} {795} (\bibinfo {year} {2013})}\BibitemShut {NoStop}%
\bibitem [{\citenamefont {Baumann}\ \emph {et~al.}(2010)\citenamefont
  {Baumann}, \citenamefont {Guerlin}, \citenamefont {Brennecke},\ and\
  \citenamefont {Esslinger}}]{Baumann2010}%
  \BibitemOpen
  \bibfield  {author} {\bibinfo {author} {\bibfnamefont {K.}~\bibnamefont
  {Baumann}}, \bibinfo {author} {\bibfnamefont {C.}~\bibnamefont {Guerlin}},
  \bibinfo {author} {\bibfnamefont {F.}~\bibnamefont {Brennecke}},\ and\
  \bibinfo {author} {\bibfnamefont {T.}~\bibnamefont {Esslinger}},\ }\href
  {https://doi.org/10.1038/nature09009} {\bibfield  {journal} {\bibinfo
  {journal} {Nature}\ }\textbf {\bibinfo {volume} {464}},\ \bibinfo {pages}
  {1301} (\bibinfo {year} {2010})}\BibitemShut {NoStop}%
\bibitem [{\citenamefont {Lieb}\ \emph {et~al.}(1961)\citenamefont {Lieb},
  \citenamefont {Schultz},\ and\ \citenamefont {Mattis}}]{Lieb1961}%
  \BibitemOpen
  \bibfield  {author} {\bibinfo {author} {\bibfnamefont {E.}~\bibnamefont
  {Lieb}}, \bibinfo {author} {\bibfnamefont {T.}~\bibnamefont {Schultz}},\ and\
  \bibinfo {author} {\bibfnamefont {D.}~\bibnamefont {Mattis}},\ }\href
  {https://doi.org/https://doi.org/10.1016/0003-4916(61)90115-4} {\bibfield
  {journal} {\bibinfo  {journal} {Annals of Physics}\ }\textbf {\bibinfo
  {volume} {16}},\ \bibinfo {pages} {407} (\bibinfo {year} {1961})}\BibitemShut
  {NoStop}%
\bibitem [{\citenamefont {Kogut}(1979)}]{Kogut1979}%
  \BibitemOpen
  \bibfield  {author} {\bibinfo {author} {\bibfnamefont {J.~B.}\ \bibnamefont
  {Kogut}},\ }\href {https://doi.org/10.1103/RevModPhys.51.659} {\bibfield
  {journal} {\bibinfo  {journal} {Rev. Mod. Phys.}\ }\textbf {\bibinfo {volume}
  {51}},\ \bibinfo {pages} {659} (\bibinfo {year} {1979})}\BibitemShut
  {NoStop}%
\bibitem [{\citenamefont {Ma}\ \emph {et~al.}(1986)\citenamefont {Ma},
  \citenamefont {Halperin},\ and\ \citenamefont {Lee}}]{Ma1986}%
  \BibitemOpen
  \bibfield  {author} {\bibinfo {author} {\bibfnamefont {M.}~\bibnamefont
  {Ma}}, \bibinfo {author} {\bibfnamefont {B.~I.}\ \bibnamefont {Halperin}},\
  and\ \bibinfo {author} {\bibfnamefont {P.~A.}\ \bibnamefont {Lee}},\ }\href
  {https://doi.org/10.1103/PhysRevB.34.3136} {\bibfield  {journal} {\bibinfo
  {journal} {Phys. Rev. B}\ }\textbf {\bibinfo {volume} {34}},\ \bibinfo
  {pages} {3136} (\bibinfo {year} {1986})}\BibitemShut {NoStop}%
\bibitem [{\citenamefont {Tobochnik}\ and\ \citenamefont
  {Chester}(1979)}]{Tobochnik1979}%
  \BibitemOpen
  \bibfield  {author} {\bibinfo {author} {\bibfnamefont {J.}~\bibnamefont
  {Tobochnik}}\ and\ \bibinfo {author} {\bibfnamefont {G.~V.}\ \bibnamefont
  {Chester}},\ }\href {https://doi.org/10.1103/PhysRevB.20.3761} {\bibfield
  {journal} {\bibinfo  {journal} {Phys. Rev. B}\ }\textbf {\bibinfo {volume}
  {20}},\ \bibinfo {pages} {3761} (\bibinfo {year} {1979})}\BibitemShut
  {NoStop}%
\bibitem [{\citenamefont {Mattis}(1984)}]{Mattis1984}%
  \BibitemOpen
  \bibfield  {author} {\bibinfo {author} {\bibfnamefont {D.~C.}\ \bibnamefont
  {Mattis}},\ }\href
  {https://doi.org/https://doi.org/10.1016/0375-9601(84)90816-8} {\bibfield
  {journal} {\bibinfo  {journal} {Physics Letters A}\ }\textbf {\bibinfo
  {volume} {104}},\ \bibinfo {pages} {357} (\bibinfo {year}
  {1984})}\BibitemShut {NoStop}%
\bibitem [{\citenamefont {Kosterlitz}\ and\ \citenamefont
  {Thouless}(1973)}]{Kosterlitz1973}%
  \BibitemOpen
  \bibfield  {author} {\bibinfo {author} {\bibfnamefont {J.~M.}\ \bibnamefont
  {Kosterlitz}}\ and\ \bibinfo {author} {\bibfnamefont {D.~J.}\ \bibnamefont
  {Thouless}},\ }\href {https://doi.org/10.1088/0022-3719/6/7/010} {\bibfield
  {journal} {\bibinfo  {journal} {Journal of Physics C: Solid State Physics}\
  }\textbf {\bibinfo {volume} {6}},\ \bibinfo {pages} {1181} (\bibinfo {year}
  {1973})}\BibitemShut {NoStop}%
\bibitem [{\citenamefont {Kosterlitz}(1974)}]{Kosterlitz1974}%
  \BibitemOpen
  \bibfield  {author} {\bibinfo {author} {\bibfnamefont {J.~M.}\ \bibnamefont
  {Kosterlitz}},\ }\href {https://doi.org/10.1088/0022-3719/7/6/005} {\bibfield
   {journal} {\bibinfo  {journal} {Journal of Physics C: Solid State Physics}\
  }\textbf {\bibinfo {volume} {7}},\ \bibinfo {pages} {1046} (\bibinfo {year}
  {1974})}\BibitemShut {NoStop}%
\bibitem [{\citenamefont {Dalla~Torre}\ \emph {et~al.}(2006)\citenamefont
  {Dalla~Torre}, \citenamefont {Berg},\ and\ \citenamefont
  {Altman}}]{DallaTorre2006}%
  \BibitemOpen
  \bibfield  {author} {\bibinfo {author} {\bibfnamefont {E.~G.}\ \bibnamefont
  {Dalla~Torre}}, \bibinfo {author} {\bibfnamefont {E.}~\bibnamefont {Berg}},\
  and\ \bibinfo {author} {\bibfnamefont {E.}~\bibnamefont {Altman}},\ }\href
  {https://doi.org/10.1103/PhysRevLett.97.260401} {\bibfield  {journal}
  {\bibinfo  {journal} {Phys. Rev. Lett.}\ }\textbf {\bibinfo {volume} {97}},\
  \bibinfo {pages} {260401} (\bibinfo {year} {2006})}\BibitemShut {NoStop}%
\bibitem [{\citenamefont {Haldane}(1983{\natexlab{a}})}]{Haldane1983}%
  \BibitemOpen
  \bibfield  {author} {\bibinfo {author} {\bibfnamefont {F.~D.~M.}\
  \bibnamefont {Haldane}},\ }\href
  {https://doi.org/10.1103/PhysRevLett.50.1153} {\bibfield  {journal} {\bibinfo
   {journal} {Phys. Rev. Lett.}\ }\textbf {\bibinfo {volume} {50}},\ \bibinfo
  {pages} {1153} (\bibinfo {year} {1983}{\natexlab{a}})}\BibitemShut {NoStop}%
\bibitem [{\citenamefont {Haldane}(1983{\natexlab{b}})}]{Haldane1983a}%
  \BibitemOpen
  \bibfield  {author} {\bibinfo {author} {\bibfnamefont {F.}~\bibnamefont
  {Haldane}},\ }\href
  {https://doi.org/https://doi.org/10.1016/0375-9601(83)90631-X} {\bibfield
  {journal} {\bibinfo  {journal} {Physics Letters A}\ }\textbf {\bibinfo
  {volume} {93}},\ \bibinfo {pages} {464 } (\bibinfo {year}
  {1983}{\natexlab{b}})}\BibitemShut {NoStop}%
\bibitem [{\citenamefont {den Nijs}\ and\ \citenamefont
  {Rommelse}(1989)}]{denNijs1989}%
  \BibitemOpen
  \bibfield  {author} {\bibinfo {author} {\bibfnamefont {M.}~\bibnamefont {den
  Nijs}}\ and\ \bibinfo {author} {\bibfnamefont {K.}~\bibnamefont {Rommelse}},\
  }\href {https://doi.org/10.1103/PhysRevB.40.4709} {\bibfield  {journal}
  {\bibinfo  {journal} {Phys. Rev. B}\ }\textbf {\bibinfo {volume} {40}},\
  \bibinfo {pages} {4709} (\bibinfo {year} {1989})}\BibitemShut {NoStop}%
\bibitem [{\citenamefont {Berg}\ \emph {et~al.}(2008)\citenamefont {Berg},
  \citenamefont {Dalla~Torre}, \citenamefont {Giamarchi},\ and\ \citenamefont
  {Altman}}]{Berg2008}%
  \BibitemOpen
  \bibfield  {author} {\bibinfo {author} {\bibfnamefont {E.}~\bibnamefont
  {Berg}}, \bibinfo {author} {\bibfnamefont {E.~G.}\ \bibnamefont
  {Dalla~Torre}}, \bibinfo {author} {\bibfnamefont {T.}~\bibnamefont
  {Giamarchi}},\ and\ \bibinfo {author} {\bibfnamefont {E.}~\bibnamefont
  {Altman}},\ }\href {https://doi.org/10.1103/PhysRevB.77.245119} {\bibfield
  {journal} {\bibinfo  {journal} {Phys. Rev. B}\ }\textbf {\bibinfo {volume}
  {77}},\ \bibinfo {pages} {245119} (\bibinfo {year} {2008})}\BibitemShut
  {NoStop}%
\bibitem [{\citenamefont {Pollock}\ and\ \citenamefont
  {Ceperley}(1987)}]{Pollock1987}%
  \BibitemOpen
  \bibfield  {author} {\bibinfo {author} {\bibfnamefont {E.~L.}\ \bibnamefont
  {Pollock}}\ and\ \bibinfo {author} {\bibfnamefont {D.~M.}\ \bibnamefont
  {Ceperley}},\ }\href {https://doi.org/10.1103/PhysRevB.36.8343} {\bibfield
  {journal} {\bibinfo  {journal} {Phys. Rev. B}\ }\textbf {\bibinfo {volume}
  {36}},\ \bibinfo {pages} {8343} (\bibinfo {year} {1987})}\BibitemShut
  {NoStop}%
\bibitem [{\citenamefont {Ceperley}(1995)}]{Ceperley1995}%
  \BibitemOpen
  \bibfield  {author} {\bibinfo {author} {\bibfnamefont {D.~M.}\ \bibnamefont
  {Ceperley}},\ }\href {https://doi.org/10.1103/RevModPhys.67.279} {\bibfield
  {journal} {\bibinfo  {journal} {Rev. Mod. Phys.}\ }\textbf {\bibinfo {volume}
  {67}},\ \bibinfo {pages} {279} (\bibinfo {year} {1995})}\BibitemShut
  {NoStop}%
\bibitem [{\citenamefont {Batrouni}\ and\ \citenamefont
  {Scalettar}(1992)}]{Batrouni1992}%
  \BibitemOpen
  \bibfield  {author} {\bibinfo {author} {\bibfnamefont {G.~G.}\ \bibnamefont
  {Batrouni}}\ and\ \bibinfo {author} {\bibfnamefont {R.~T.}\ \bibnamefont
  {Scalettar}},\ }\href {https://doi.org/10.1103/PhysRevB.46.9051} {\bibfield
  {journal} {\bibinfo  {journal} {Phys. Rev. B}\ }\textbf {\bibinfo {volume}
  {46}},\ \bibinfo {pages} {9051} (\bibinfo {year} {1992})}\BibitemShut
  {NoStop}%
\bibitem [{\citenamefont {Batrouni}\ \emph {et~al.}(1995)\citenamefont
  {Batrouni}, \citenamefont {Scalettar}, \citenamefont {Zimanyi},\ and\
  \citenamefont {Kampf}}]{Batrouni1995}%
  \BibitemOpen
  \bibfield  {author} {\bibinfo {author} {\bibfnamefont {G.~G.}\ \bibnamefont
  {Batrouni}}, \bibinfo {author} {\bibfnamefont {R.~T.}\ \bibnamefont
  {Scalettar}}, \bibinfo {author} {\bibfnamefont {G.~T.}\ \bibnamefont
  {Zimanyi}},\ and\ \bibinfo {author} {\bibfnamefont {A.~P.}\ \bibnamefont
  {Kampf}},\ }\href {https://doi.org/10.1103/PhysRevLett.74.2527} {\bibfield
  {journal} {\bibinfo  {journal} {Phys. Rev. Lett.}\ }\textbf {\bibinfo
  {volume} {74}},\ \bibinfo {pages} {2527} (\bibinfo {year}
  {1995})}\BibitemShut {NoStop}%
\bibitem [{\citenamefont {Prokof'ev}\ \emph
  {et~al.}(1998{\natexlab{a}})\citenamefont {Prokof'ev}, \citenamefont
  {Svistunov},\ and\ \citenamefont {Tupitsyn}}]{Prokofev1998}%
  \BibitemOpen
  \bibfield  {author} {\bibinfo {author} {\bibfnamefont {N.}~\bibnamefont
  {Prokof'ev}}, \bibinfo {author} {\bibfnamefont {B.}~\bibnamefont
  {Svistunov}},\ and\ \bibinfo {author} {\bibfnamefont {I.}~\bibnamefont
  {Tupitsyn}},\ }\href
  {https://doi.org/http://dx.doi.org/10.1016/S0375-9601(97)00957-2} {\bibfield
  {journal} {\bibinfo  {journal} {Physics Letters A}\ }\textbf {\bibinfo
  {volume} {238}},\ \bibinfo {pages} {253 } (\bibinfo {year}
  {1998}{\natexlab{a}})}\BibitemShut {NoStop}%
\bibitem [{\citenamefont {Prokof'ev}\ \emph
  {et~al.}(1998{\natexlab{b}})\citenamefont {Prokof'ev}, \citenamefont
  {Svistunov},\ and\ \citenamefont {Tupitsyn}}]{Prokofev1998a}%
  \BibitemOpen
  \bibfield  {author} {\bibinfo {author} {\bibfnamefont {N.~V.}\ \bibnamefont
  {Prokof'ev}}, \bibinfo {author} {\bibfnamefont {B.~V.}\ \bibnamefont
  {Svistunov}},\ and\ \bibinfo {author} {\bibfnamefont {I.~S.}\ \bibnamefont
  {Tupitsyn}},\ }\href {https://doi.org/10.1134/1.558661} {\bibfield  {journal}
  {\bibinfo  {journal} {Journal of Experimental and Theoretical Physics}\
  }\textbf {\bibinfo {volume} {87}},\ \bibinfo {pages} {310} (\bibinfo {year}
  {1998}{\natexlab{b}})}\BibitemShut {NoStop}%
\bibitem [{\citenamefont {van Oosten}\ \emph {et~al.}(2001)\citenamefont {van
  Oosten}, \citenamefont {van~der Straten},\ and\ \citenamefont
  {Stoof}}]{Oosten2001}%
  \BibitemOpen
  \bibfield  {author} {\bibinfo {author} {\bibfnamefont {D.}~\bibnamefont {van
  Oosten}}, \bibinfo {author} {\bibfnamefont {P.}~\bibnamefont {van~der
  Straten}},\ and\ \bibinfo {author} {\bibfnamefont {H.~T.~C.}\ \bibnamefont
  {Stoof}},\ }\href {https://doi.org/10.1103/PhysRevA.63.053601} {\bibfield
  {journal} {\bibinfo  {journal} {Phys. Rev. A}\ }\textbf {\bibinfo {volume}
  {63}},\ \bibinfo {pages} {053601} (\bibinfo {year} {2001})}\BibitemShut
  {NoStop}%
\bibitem [{\citenamefont {K\"uhner}\ and\ \citenamefont
  {Monien}(1998)}]{Kuehner1998}%
  \BibitemOpen
  \bibfield  {author} {\bibinfo {author} {\bibfnamefont {T.~D.}\ \bibnamefont
  {K\"uhner}}\ and\ \bibinfo {author} {\bibfnamefont {H.}~\bibnamefont
  {Monien}},\ }\href {https://doi.org/10.1103/PhysRevB.58.R14741} {\bibfield
  {journal} {\bibinfo  {journal} {Phys. Rev. B}\ }\textbf {\bibinfo {volume}
  {58}},\ \bibinfo {pages} {R14741} (\bibinfo {year} {1998})}\BibitemShut
  {NoStop}%
\bibitem [{\citenamefont {Batrouni}\ \emph {et~al.}(2013)\citenamefont
  {Batrouni}, \citenamefont {Scalettar}, \citenamefont {Rousseau},\ and\
  \citenamefont {Gr\'emaud}}]{Batrouni2013}%
  \BibitemOpen
  \bibfield  {author} {\bibinfo {author} {\bibfnamefont {G.~G.}\ \bibnamefont
  {Batrouni}}, \bibinfo {author} {\bibfnamefont {R.~T.}\ \bibnamefont
  {Scalettar}}, \bibinfo {author} {\bibfnamefont {V.~G.}\ \bibnamefont
  {Rousseau}},\ and\ \bibinfo {author} {\bibfnamefont {B.}~\bibnamefont
  {Gr\'emaud}},\ }\href {https://doi.org/10.1103/PhysRevLett.110.265303}
  {\bibfield  {journal} {\bibinfo  {journal} {Phys. Rev. Lett.}\ }\textbf
  {\bibinfo {volume} {110}},\ \bibinfo {pages} {265303} (\bibinfo {year}
  {2013})}\BibitemShut {NoStop}%
\bibitem [{\citenamefont {Ohgoe}\ \emph {et~al.}(2012)\citenamefont {Ohgoe},
  \citenamefont {Suzuki},\ and\ \citenamefont {Kawashima}}]{Ohgoe2012}%
  \BibitemOpen
  \bibfield  {author} {\bibinfo {author} {\bibfnamefont {T.}~\bibnamefont
  {Ohgoe}}, \bibinfo {author} {\bibfnamefont {T.}~\bibnamefont {Suzuki}},\ and\
  \bibinfo {author} {\bibfnamefont {N.}~\bibnamefont {Kawashima}},\ }\href
  {https://doi.org/10.1103/PhysRevB.86.054520} {\bibfield  {journal} {\bibinfo
  {journal} {Phys. Rev. B}\ }\textbf {\bibinfo {volume} {86}},\ \bibinfo
  {pages} {054520} (\bibinfo {year} {2012})}\BibitemShut {NoStop}%
\bibitem [{\citenamefont {Sengupta}\ \emph {et~al.}(2005)\citenamefont
  {Sengupta}, \citenamefont {Pryadko}, \citenamefont {Alet}, \citenamefont
  {Troyer},\ and\ \citenamefont {Schmid}}]{Sengupta2005}%
  \BibitemOpen
  \bibfield  {author} {\bibinfo {author} {\bibfnamefont {P.}~\bibnamefont
  {Sengupta}}, \bibinfo {author} {\bibfnamefont {L.~P.}\ \bibnamefont
  {Pryadko}}, \bibinfo {author} {\bibfnamefont {F.}~\bibnamefont {Alet}},
  \bibinfo {author} {\bibfnamefont {M.}~\bibnamefont {Troyer}},\ and\ \bibinfo
  {author} {\bibfnamefont {G.}~\bibnamefont {Schmid}},\ }\href
  {https://doi.org/10.1103/PhysRevLett.94.207202} {\bibfield  {journal}
  {\bibinfo  {journal} {Phys. Rev. Lett.}\ }\textbf {\bibinfo {volume} {94}},\
  \bibinfo {pages} {207202} (\bibinfo {year} {2005})}\BibitemShut {NoStop}%
\bibitem [{\citenamefont {Batrouni}\ \emph {et~al.}(2006)\citenamefont
  {Batrouni}, \citenamefont {H\'ebert},\ and\ \citenamefont
  {Scalettar}}]{Batrouni2006}%
  \BibitemOpen
  \bibfield  {author} {\bibinfo {author} {\bibfnamefont {G.~G.}\ \bibnamefont
  {Batrouni}}, \bibinfo {author} {\bibfnamefont {F.}~\bibnamefont {H\'ebert}},\
  and\ \bibinfo {author} {\bibfnamefont {R.~T.}\ \bibnamefont {Scalettar}},\
  }\href {https://doi.org/10.1103/PhysRevLett.97.087209} {\bibfield  {journal}
  {\bibinfo  {journal} {Phys. Rev. Lett.}\ }\textbf {\bibinfo {volume} {97}},\
  \bibinfo {pages} {087209} (\bibinfo {year} {2006})}\BibitemShut {NoStop}%
\bibitem [{\citenamefont {Iskin}(2011)}]{Iskin2011}%
  \BibitemOpen
  \bibfield  {author} {\bibinfo {author} {\bibfnamefont {M.}~\bibnamefont
  {Iskin}},\ }\href {https://doi.org/10.1103/PhysRevA.83.051606} {\bibfield
  {journal} {\bibinfo  {journal} {Phys. Rev. A}\ }\textbf {\bibinfo {volume}
  {83}},\ \bibinfo {pages} {051606(R)} (\bibinfo {year} {2011})}\BibitemShut
  {NoStop}%
\bibitem [{\citenamefont {Rossini}\ and\ \citenamefont
  {Fazio}(2012)}]{Rossini2012}%
  \BibitemOpen
  \bibfield  {author} {\bibinfo {author} {\bibfnamefont {D.}~\bibnamefont
  {Rossini}}\ and\ \bibinfo {author} {\bibfnamefont {R.}~\bibnamefont
  {Fazio}},\ }\href {https://doi.org/10.1088/1367-2630/14/6/065012} {\bibfield
  {journal} {\bibinfo  {journal} {New Journal of Physics}\ }\textbf {\bibinfo
  {volume} {14}},\ \bibinfo {pages} {065012} (\bibinfo {year}
  {2012})}\BibitemShut {NoStop}%
\bibitem [{\citenamefont {Kawaki}\ \emph {et~al.}(2017)\citenamefont {Kawaki},
  \citenamefont {Kuno},\ and\ \citenamefont {Ichinose}}]{Kawaki2017}%
  \BibitemOpen
  \bibfield  {author} {\bibinfo {author} {\bibfnamefont {K.}~\bibnamefont
  {Kawaki}}, \bibinfo {author} {\bibfnamefont {Y.}~\bibnamefont {Kuno}},\ and\
  \bibinfo {author} {\bibfnamefont {I.}~\bibnamefont {Ichinose}},\ }\href
  {https://doi.org/10.1103/PhysRevB.95.195101} {\bibfield  {journal} {\bibinfo
  {journal} {Phys. Rev. B}\ }\textbf {\bibinfo {volume} {95}},\ \bibinfo
  {pages} {195101} (\bibinfo {year} {2017})}\BibitemShut {NoStop}%
\bibitem [{\citenamefont {Schmid}\ and\ \citenamefont
  {Troyer}(2004)}]{Schmid2004}%
  \BibitemOpen
  \bibfield  {author} {\bibinfo {author} {\bibfnamefont {G.}~\bibnamefont
  {Schmid}}\ and\ \bibinfo {author} {\bibfnamefont {M.}~\bibnamefont
  {Troyer}},\ }\href {https://doi.org/10.1103/PhysRevLett.93.067003} {\bibfield
   {journal} {\bibinfo  {journal} {Phys. Rev. Lett.}\ }\textbf {\bibinfo
  {volume} {93}},\ \bibinfo {pages} {067003} (\bibinfo {year}
  {2004})}\BibitemShut {NoStop}%
\bibitem [{\citenamefont {Chen}\ \emph {et~al.}(2008)\citenamefont {Chen},
  \citenamefont {Melko}, \citenamefont {Wessel},\ and\ \citenamefont
  {Kao}}]{Chen2008}%
  \BibitemOpen
  \bibfield  {author} {\bibinfo {author} {\bibfnamefont {Y.-C.}\ \bibnamefont
  {Chen}}, \bibinfo {author} {\bibfnamefont {R.~G.}\ \bibnamefont {Melko}},
  \bibinfo {author} {\bibfnamefont {S.}~\bibnamefont {Wessel}},\ and\ \bibinfo
  {author} {\bibfnamefont {Y.-J.}\ \bibnamefont {Kao}},\ }\href
  {https://doi.org/10.1103/PhysRevB.77.014524} {\bibfield  {journal} {\bibinfo
  {journal} {Phys. Rev. B}\ }\textbf {\bibinfo {volume} {77}},\ \bibinfo
  {pages} {014524} (\bibinfo {year} {2008})}\BibitemShut {NoStop}%
\bibitem [{\citenamefont {Dogra}\ \emph {et~al.}(2016)\citenamefont {Dogra},
  \citenamefont {Brennecke}, \citenamefont {Huber},\ and\ \citenamefont
  {Donner}}]{Dogra2016}%
  \BibitemOpen
  \bibfield  {author} {\bibinfo {author} {\bibfnamefont {N.}~\bibnamefont
  {Dogra}}, \bibinfo {author} {\bibfnamefont {F.}~\bibnamefont {Brennecke}},
  \bibinfo {author} {\bibfnamefont {S.~D.}\ \bibnamefont {Huber}},\ and\
  \bibinfo {author} {\bibfnamefont {T.}~\bibnamefont {Donner}},\ }\href
  {https://doi.org/10.1103/PhysRevA.94.023632} {\bibfield  {journal} {\bibinfo
  {journal} {Phys. Rev. A}\ }\textbf {\bibinfo {volume} {94}},\ \bibinfo
  {pages} {023632} (\bibinfo {year} {2016})}\BibitemShut {NoStop}%
\bibitem [{\citenamefont {Flottat}\ \emph {et~al.}(2017)\citenamefont
  {Flottat}, \citenamefont {de~Parny}, \citenamefont {H\'ebert}, \citenamefont
  {Rousseau},\ and\ \citenamefont {Batrouni}}]{Flottat2017}%
  \BibitemOpen
  \bibfield  {author} {\bibinfo {author} {\bibfnamefont {T.}~\bibnamefont
  {Flottat}}, \bibinfo {author} {\bibfnamefont {L.~d.~F.}\ \bibnamefont
  {de~Parny}}, \bibinfo {author} {\bibfnamefont {F.}~\bibnamefont {H\'ebert}},
  \bibinfo {author} {\bibfnamefont {V.~G.}\ \bibnamefont {Rousseau}},\ and\
  \bibinfo {author} {\bibfnamefont {G.~G.}\ \bibnamefont {Batrouni}},\ }\href
  {https://doi.org/10.1103/PhysRevB.95.144501} {\bibfield  {journal} {\bibinfo
  {journal} {Phys. Rev. B}\ }\textbf {\bibinfo {volume} {95}},\ \bibinfo
  {pages} {144501} (\bibinfo {year} {2017})}\BibitemShut {NoStop}%
\bibitem [{\citenamefont {Hruby}\ \emph {et~al.}(2018)\citenamefont {Hruby},
  \citenamefont {Dogra}, \citenamefont {Landini}, \citenamefont {Donner},\ and\
  \citenamefont {Esslinger}}]{Hruby2018}%
  \BibitemOpen
  \bibfield  {author} {\bibinfo {author} {\bibfnamefont {L.}~\bibnamefont
  {Hruby}}, \bibinfo {author} {\bibfnamefont {N.}~\bibnamefont {Dogra}},
  \bibinfo {author} {\bibfnamefont {M.}~\bibnamefont {Landini}}, \bibinfo
  {author} {\bibfnamefont {T.}~\bibnamefont {Donner}},\ and\ \bibinfo {author}
  {\bibfnamefont {T.}~\bibnamefont {Esslinger}},\ }\href
  {https://doi.org/10.1073/pnas.1720415115} {\bibfield  {journal} {\bibinfo
  {journal} {Proceedings of the National Academy of Sciences}\ }\textbf
  {\bibinfo {volume} {115}},\ \bibinfo {pages} {3279} (\bibinfo {year}
  {2018})},\ \Eprint
  {https://arxiv.org/abs/https://www.pnas.org/content/115/13/3279.full.pdf}
  {https://www.pnas.org/content/115/13/3279.full.pdf} \BibitemShut {NoStop}%
\bibitem [{\citenamefont {{Bogner, Benjamin}}\ \emph
  {et~al.}(2019)\citenamefont {{Bogner, Benjamin}}, \citenamefont {{De
  Daniloff, Cl\'ement}},\ and\ \citenamefont {{Rieger, Heiko}}}]{Bogner2019}%
  \BibitemOpen
  \bibfield  {author} {\bibinfo {author} {\bibnamefont {{Bogner, Benjamin}}},
  \bibinfo {author} {\bibnamefont {{De Daniloff, Cl\'ement}}},\ and\ \bibinfo
  {author} {\bibnamefont {{Rieger, Heiko}}},\ }\href
  {https://doi.org/10.1140/epjb/e2019-100017-8} {\bibfield  {journal} {\bibinfo
   {journal} {Eur. Phys. J. B}\ }\textbf {\bibinfo {volume} {92}},\ \bibinfo
  {pages} {111} (\bibinfo {year} {2019})}\BibitemShut {NoStop}%
\bibitem [{\citenamefont {Sicks}\ and\ \citenamefont
  {Rieger}(2020)}]{Sicks2020}%
  \BibitemOpen
  \bibfield  {author} {\bibinfo {author} {\bibfnamefont {J.}~\bibnamefont
  {Sicks}}\ and\ \bibinfo {author} {\bibfnamefont {H.}~\bibnamefont {Rieger}},\
  }\href {https://doi.org/10.1140/epjb/e2020-10109-3} {\bibfield  {journal}
  {\bibinfo  {journal} {Eur. Phys. J. B}\ }\textbf {\bibinfo {volume} {93}},\
  \bibinfo {pages} {104} (\bibinfo {year} {2020})}\BibitemShut {NoStop}%
\bibitem [{\citenamefont {Gurarie}\ \emph {et~al.}(2009)\citenamefont
  {Gurarie}, \citenamefont {Pollet}, \citenamefont {Prokof'ev}, \citenamefont
  {Svistunov},\ and\ \citenamefont {Troyer}}]{Gurarie2009}%
  \BibitemOpen
  \bibfield  {author} {\bibinfo {author} {\bibfnamefont {V.}~\bibnamefont
  {Gurarie}}, \bibinfo {author} {\bibfnamefont {L.}~\bibnamefont {Pollet}},
  \bibinfo {author} {\bibfnamefont {N.~V.}\ \bibnamefont {Prokof'ev}}, \bibinfo
  {author} {\bibfnamefont {B.~V.}\ \bibnamefont {Svistunov}},\ and\ \bibinfo
  {author} {\bibfnamefont {M.}~\bibnamefont {Troyer}},\ }\href
  {https://doi.org/10.1103/PhysRevB.80.214519} {\bibfield  {journal} {\bibinfo
  {journal} {Phys. Rev. B}\ }\textbf {\bibinfo {volume} {80}},\ \bibinfo
  {pages} {214519} (\bibinfo {year} {2009})}\BibitemShut {NoStop}%
\bibitem [{\citenamefont {Niederle}\ \emph {et~al.}(2016)\citenamefont
  {Niederle}, \citenamefont {Morigi},\ and\ \citenamefont
  {Rieger}}]{Niederle2016}%
  \BibitemOpen
  \bibfield  {author} {\bibinfo {author} {\bibfnamefont {A.~E.}\ \bibnamefont
  {Niederle}}, \bibinfo {author} {\bibfnamefont {G.}~\bibnamefont {Morigi}},\
  and\ \bibinfo {author} {\bibfnamefont {H.}~\bibnamefont {Rieger}},\ }\href
  {https://doi.org/10.1103/PhysRevA.94.033607} {\bibfield  {journal} {\bibinfo
  {journal} {Phys. Rev. A}\ }\textbf {\bibinfo {volume} {94}},\ \bibinfo
  {pages} {033607} (\bibinfo {year} {2016})}\BibitemShut {NoStop}%
\bibitem [{\citenamefont {Sebby-Strabley}\ \emph {et~al.}(2006)\citenamefont
  {Sebby-Strabley}, \citenamefont {Anderlini}, \citenamefont {Jessen},\ and\
  \citenamefont {Porto}}]{SebbyStrabley2006}%
  \BibitemOpen
  \bibfield  {author} {\bibinfo {author} {\bibfnamefont {J.}~\bibnamefont
  {Sebby-Strabley}}, \bibinfo {author} {\bibfnamefont {M.}~\bibnamefont
  {Anderlini}}, \bibinfo {author} {\bibfnamefont {P.~S.}\ \bibnamefont
  {Jessen}},\ and\ \bibinfo {author} {\bibfnamefont {J.~V.}\ \bibnamefont
  {Porto}},\ }\href {https://doi.org/10.1103/PhysRevA.73.033605} {\bibfield
  {journal} {\bibinfo  {journal} {Phys. Rev. A}\ }\textbf {\bibinfo {volume}
  {73}},\ \bibinfo {pages} {033605} (\bibinfo {year} {2006})}\BibitemShut
  {NoStop}%
\bibitem [{\citenamefont {F{\"o}lling}\ \emph {et~al.}(2007)\citenamefont
  {F{\"o}lling}, \citenamefont {Trotzky}, \citenamefont {Cheinet},
  \citenamefont {Feld}, \citenamefont {Saers}, \citenamefont {Widera},
  \citenamefont {M{\"u}ller},\ and\ \citenamefont {Bloch}}]{Foelling2007}%
  \BibitemOpen
  \bibfield  {author} {\bibinfo {author} {\bibfnamefont {S.}~\bibnamefont
  {F{\"o}lling}}, \bibinfo {author} {\bibfnamefont {S.}~\bibnamefont
  {Trotzky}}, \bibinfo {author} {\bibfnamefont {P.}~\bibnamefont {Cheinet}},
  \bibinfo {author} {\bibfnamefont {M.}~\bibnamefont {Feld}}, \bibinfo {author}
  {\bibfnamefont {R.}~\bibnamefont {Saers}}, \bibinfo {author} {\bibfnamefont
  {A.}~\bibnamefont {Widera}}, \bibinfo {author} {\bibfnamefont
  {T.}~\bibnamefont {M{\"u}ller}},\ and\ \bibinfo {author} {\bibfnamefont
  {I.}~\bibnamefont {Bloch}},\ }\href {https://doi.org/10.1038/nature06112}
  {\bibfield  {journal} {\bibinfo  {journal} {Nature}\ }\textbf {\bibinfo
  {volume} {448}},\ \bibinfo {pages} {1029} (\bibinfo {year}
  {2007})}\BibitemShut {NoStop}%
\bibitem [{\citenamefont {Trotzky}\ \emph {et~al.}(2008)\citenamefont
  {Trotzky}, \citenamefont {Cheinet}, \citenamefont {Fölling}, \citenamefont
  {Feld}, \citenamefont {Schnorrberger}, \citenamefont {Rey}, \citenamefont
  {Polkovnikov}, \citenamefont {Demler}, \citenamefont {Lukin},\ and\
  \citenamefont {Bloch}}]{Trotzky2008}%
  \BibitemOpen
  \bibfield  {author} {\bibinfo {author} {\bibfnamefont {S.}~\bibnamefont
  {Trotzky}}, \bibinfo {author} {\bibfnamefont {P.}~\bibnamefont {Cheinet}},
  \bibinfo {author} {\bibfnamefont {S.}~\bibnamefont {Fölling}}, \bibinfo
  {author} {\bibfnamefont {M.}~\bibnamefont {Feld}}, \bibinfo {author}
  {\bibfnamefont {U.}~\bibnamefont {Schnorrberger}}, \bibinfo {author}
  {\bibfnamefont {A.~M.}\ \bibnamefont {Rey}}, \bibinfo {author} {\bibfnamefont
  {A.}~\bibnamefont {Polkovnikov}}, \bibinfo {author} {\bibfnamefont {E.~A.}\
  \bibnamefont {Demler}}, \bibinfo {author} {\bibfnamefont {M.~D.}\
  \bibnamefont {Lukin}},\ and\ \bibinfo {author} {\bibfnamefont
  {I.}~\bibnamefont {Bloch}},\ }\href {https://doi.org/10.1126/science.1150841}
  {\bibfield  {journal} {\bibinfo  {journal} {Science}\ }\textbf {\bibinfo
  {volume} {319}},\ \bibinfo {pages} {295} (\bibinfo {year} {2008})},\ \Eprint
  {https://arxiv.org/abs/https://www.science.org/doi/pdf/10.1126/science.1150841}
  {https://www.science.org/doi/pdf/10.1126/science.1150841} \BibitemShut
  {NoStop}%
\bibitem [{\citenamefont {Barmettler}\ \emph {et~al.}(2008)\citenamefont
  {Barmettler}, \citenamefont {Rey}, \citenamefont {Demler}, \citenamefont
  {Lukin}, \citenamefont {Bloch},\ and\ \citenamefont
  {Gritsev}}]{Barmettler2008}%
  \BibitemOpen
  \bibfield  {author} {\bibinfo {author} {\bibfnamefont {P.}~\bibnamefont
  {Barmettler}}, \bibinfo {author} {\bibfnamefont {A.~M.}\ \bibnamefont {Rey}},
  \bibinfo {author} {\bibfnamefont {E.}~\bibnamefont {Demler}}, \bibinfo
  {author} {\bibfnamefont {M.~D.}\ \bibnamefont {Lukin}}, \bibinfo {author}
  {\bibfnamefont {I.}~\bibnamefont {Bloch}},\ and\ \bibinfo {author}
  {\bibfnamefont {V.}~\bibnamefont {Gritsev}},\ }\href
  {https://doi.org/10.1103/PhysRevA.78.012330} {\bibfield  {journal} {\bibinfo
  {journal} {Phys. Rev. A}\ }\textbf {\bibinfo {volume} {78}},\ \bibinfo
  {pages} {012330} (\bibinfo {year} {2008})}\BibitemShut {NoStop}%
\bibitem [{\citenamefont {Yin}\ \emph {et~al.}(2015)\citenamefont {Yin},
  \citenamefont {Cao},\ and\ \citenamefont {Schmelcher}}]{Yin2015}%
  \BibitemOpen
  \bibfield  {author} {\bibinfo {author} {\bibfnamefont {X.}~\bibnamefont
  {Yin}}, \bibinfo {author} {\bibfnamefont {L.}~\bibnamefont {Cao}},\ and\
  \bibinfo {author} {\bibfnamefont {P.}~\bibnamefont {Schmelcher}},\ }\href
  {https://doi.org/10.1209/0295-5075/110/26004} {\bibfield  {journal} {\bibinfo
   {journal} {Europhysics Letters}\ }\textbf {\bibinfo {volume} {110}},\
  \bibinfo {pages} {26004} (\bibinfo {year} {2015})}\BibitemShut {NoStop}%
\bibitem [{\citenamefont {Volosniev}\ \emph {et~al.}(2015)\citenamefont
  {Volosniev}, \citenamefont {Petrosyan}, \citenamefont {Valiente},
  \citenamefont {Fedorov}, \citenamefont {Jensen},\ and\ \citenamefont
  {Zinner}}]{Volosniev2015}%
  \BibitemOpen
  \bibfield  {author} {\bibinfo {author} {\bibfnamefont {A.~G.}\ \bibnamefont
  {Volosniev}}, \bibinfo {author} {\bibfnamefont {D.}~\bibnamefont
  {Petrosyan}}, \bibinfo {author} {\bibfnamefont {M.}~\bibnamefont {Valiente}},
  \bibinfo {author} {\bibfnamefont {D.~V.}\ \bibnamefont {Fedorov}}, \bibinfo
  {author} {\bibfnamefont {A.~S.}\ \bibnamefont {Jensen}},\ and\ \bibinfo
  {author} {\bibfnamefont {N.~T.}\ \bibnamefont {Zinner}},\ }\href
  {https://doi.org/10.1103/PhysRevA.91.023620} {\bibfield  {journal} {\bibinfo
  {journal} {Phys. Rev. A}\ }\textbf {\bibinfo {volume} {91}},\ \bibinfo
  {pages} {023620} (\bibinfo {year} {2015})}\BibitemShut {NoStop}%
\bibitem [{\citenamefont {Mondal}\ \emph {et~al.}(2019)\citenamefont {Mondal},
  \citenamefont {Greschner},\ and\ \citenamefont {Mishra}}]{Mondal2019}%
  \BibitemOpen
  \bibfield  {author} {\bibinfo {author} {\bibfnamefont {S.}~\bibnamefont
  {Mondal}}, \bibinfo {author} {\bibfnamefont {S.}~\bibnamefont {Greschner}},\
  and\ \bibinfo {author} {\bibfnamefont {T.}~\bibnamefont {Mishra}},\ }\href
  {https://doi.org/10.1103/PhysRevA.100.013627} {\bibfield  {journal} {\bibinfo
   {journal} {Phys. Rev. A}\ }\textbf {\bibinfo {volume} {100}},\ \bibinfo
  {pages} {013627} (\bibinfo {year} {2019})}\BibitemShut {NoStop}%
\bibitem [{\citenamefont {Peil}\ \emph {et~al.}(2003)\citenamefont {Peil},
  \citenamefont {Porto}, \citenamefont {Laburthe~Tolra}, \citenamefont
  {Obrecht}, \citenamefont {King}, \citenamefont {Subbotin}, \citenamefont
  {Rolston},\ and\ \citenamefont {Phillips}}]{Peil2003}%
  \BibitemOpen
  \bibfield  {author} {\bibinfo {author} {\bibfnamefont {S.}~\bibnamefont
  {Peil}}, \bibinfo {author} {\bibfnamefont {J.~V.}\ \bibnamefont {Porto}},
  \bibinfo {author} {\bibfnamefont {B.}~\bibnamefont {Laburthe~Tolra}},
  \bibinfo {author} {\bibfnamefont {J.~M.}\ \bibnamefont {Obrecht}}, \bibinfo
  {author} {\bibfnamefont {B.~E.}\ \bibnamefont {King}}, \bibinfo {author}
  {\bibfnamefont {M.}~\bibnamefont {Subbotin}}, \bibinfo {author}
  {\bibfnamefont {S.~L.}\ \bibnamefont {Rolston}},\ and\ \bibinfo {author}
  {\bibfnamefont {W.~D.}\ \bibnamefont {Phillips}},\ }\href
  {https://doi.org/10.1103/PhysRevA.67.051603} {\bibfield  {journal} {\bibinfo
  {journal} {Phys. Rev. A}\ }\textbf {\bibinfo {volume} {67}},\ \bibinfo
  {pages} {051603(R)} (\bibinfo {year} {2003})}\BibitemShut {NoStop}%
\bibitem [{\citenamefont {Brennen}\ \emph {et~al.}(1999)\citenamefont
  {Brennen}, \citenamefont {Caves}, \citenamefont {Jessen},\ and\ \citenamefont
  {Deutsch}}]{Brennen1999}%
  \BibitemOpen
  \bibfield  {author} {\bibinfo {author} {\bibfnamefont {G.~K.}\ \bibnamefont
  {Brennen}}, \bibinfo {author} {\bibfnamefont {C.~M.}\ \bibnamefont {Caves}},
  \bibinfo {author} {\bibfnamefont {P.~S.}\ \bibnamefont {Jessen}},\ and\
  \bibinfo {author} {\bibfnamefont {I.~H.}\ \bibnamefont {Deutsch}},\ }\href
  {https://doi.org/10.1103/PhysRevLett.82.1060} {\bibfield  {journal} {\bibinfo
   {journal} {Phys. Rev. Lett.}\ }\textbf {\bibinfo {volume} {82}},\ \bibinfo
  {pages} {1060} (\bibinfo {year} {1999})}\BibitemShut {NoStop}%
\bibitem [{\citenamefont {Yang}\ \emph {et~al.}(2020)\citenamefont {Yang},
  \citenamefont {Sun}, \citenamefont {Huang}, \citenamefont {Wang},
  \citenamefont {Deng}, \citenamefont {Dai}, \citenamefont {Yuan},\ and\
  \citenamefont {Pan}}]{Yang2020}%
  \BibitemOpen
  \bibfield  {author} {\bibinfo {author} {\bibfnamefont {B.}~\bibnamefont
  {Yang}}, \bibinfo {author} {\bibfnamefont {H.}~\bibnamefont {Sun}}, \bibinfo
  {author} {\bibfnamefont {C.-J.}\ \bibnamefont {Huang}}, \bibinfo {author}
  {\bibfnamefont {H.-Y.}\ \bibnamefont {Wang}}, \bibinfo {author}
  {\bibfnamefont {Y.}~\bibnamefont {Deng}}, \bibinfo {author} {\bibfnamefont
  {H.-N.}\ \bibnamefont {Dai}}, \bibinfo {author} {\bibfnamefont {Z.-S.}\
  \bibnamefont {Yuan}},\ and\ \bibinfo {author} {\bibfnamefont {J.-W.}\
  \bibnamefont {Pan}},\ }\href {https://doi.org/10.1126/science.aaz6801}
  {\bibfield  {journal} {\bibinfo  {journal} {Science}\ }\textbf {\bibinfo
  {volume} {369}},\ \bibinfo {pages} {550} (\bibinfo {year} {2020})},\ \Eprint
  {https://arxiv.org/abs/https://www.science.org/doi/pdf/10.1126/science.aaz6801}
  {https://www.science.org/doi/pdf/10.1126/science.aaz6801} \BibitemShut
  {NoStop}%
\bibitem [{\citenamefont {Lee}\ \emph {et~al.}(2007)\citenamefont {Lee},
  \citenamefont {Anderlini}, \citenamefont {Brown}, \citenamefont
  {Sebby-Strabley}, \citenamefont {Phillips},\ and\ \citenamefont
  {Porto}}]{Lee2007}%
  \BibitemOpen
  \bibfield  {author} {\bibinfo {author} {\bibfnamefont {P.~J.}\ \bibnamefont
  {Lee}}, \bibinfo {author} {\bibfnamefont {M.}~\bibnamefont {Anderlini}},
  \bibinfo {author} {\bibfnamefont {B.~L.}\ \bibnamefont {Brown}}, \bibinfo
  {author} {\bibfnamefont {J.}~\bibnamefont {Sebby-Strabley}}, \bibinfo
  {author} {\bibfnamefont {W.~D.}\ \bibnamefont {Phillips}},\ and\ \bibinfo
  {author} {\bibfnamefont {J.~V.}\ \bibnamefont {Porto}},\ }\href
  {https://doi.org/10.1103/PhysRevLett.99.020402} {\bibfield  {journal}
  {\bibinfo  {journal} {Phys. Rev. Lett.}\ }\textbf {\bibinfo {volume} {99}},\
  \bibinfo {pages} {020402} (\bibinfo {year} {2007})}\BibitemShut {NoStop}%
\bibitem [{\citenamefont {Su}\ \emph {et~al.}(1979)\citenamefont {Su},
  \citenamefont {Schrieffer},\ and\ \citenamefont {Heeger}}]{Su1979}%
  \BibitemOpen
  \bibfield  {author} {\bibinfo {author} {\bibfnamefont {W.~P.}\ \bibnamefont
  {Su}}, \bibinfo {author} {\bibfnamefont {J.~R.}\ \bibnamefont {Schrieffer}},\
  and\ \bibinfo {author} {\bibfnamefont {A.~J.}\ \bibnamefont {Heeger}},\
  }\href {https://doi.org/10.1103/PhysRevLett.42.1698} {\bibfield  {journal}
  {\bibinfo  {journal} {Phys. Rev. Lett.}\ }\textbf {\bibinfo {volume} {42}},\
  \bibinfo {pages} {1698} (\bibinfo {year} {1979})}\BibitemShut {NoStop}%
\bibitem [{\citenamefont {Grusdt}\ \emph {et~al.}(2013)\citenamefont {Grusdt},
  \citenamefont {H\"oning},\ and\ \citenamefont {Fleischhauer}}]{Grusdt2013}%
  \BibitemOpen
  \bibfield  {author} {\bibinfo {author} {\bibfnamefont {F.}~\bibnamefont
  {Grusdt}}, \bibinfo {author} {\bibfnamefont {M.}~\bibnamefont {H\"oning}},\
  and\ \bibinfo {author} {\bibfnamefont {M.}~\bibnamefont {Fleischhauer}},\
  }\href {https://doi.org/10.1103/PhysRevLett.110.260405} {\bibfield  {journal}
  {\bibinfo  {journal} {Phys. Rev. Lett.}\ }\textbf {\bibinfo {volume} {110}},\
  \bibinfo {pages} {260405} (\bibinfo {year} {2013})}\BibitemShut {NoStop}%
\bibitem [{\citenamefont {Hayashi}\ \emph {et~al.}(2022)\citenamefont
  {Hayashi}, \citenamefont {Mondal}, \citenamefont {Mishra},\ and\
  \citenamefont {Das}}]{Hayashi2022}%
  \BibitemOpen
  \bibfield  {author} {\bibinfo {author} {\bibfnamefont {A.}~\bibnamefont
  {Hayashi}}, \bibinfo {author} {\bibfnamefont {S.}~\bibnamefont {Mondal}},
  \bibinfo {author} {\bibfnamefont {T.}~\bibnamefont {Mishra}},\ and\ \bibinfo
  {author} {\bibfnamefont {B.~P.}\ \bibnamefont {Das}},\ }\href
  {https://doi.org/10.1103/PhysRevA.106.013313} {\bibfield  {journal} {\bibinfo
   {journal} {Phys. Rev. A}\ }\textbf {\bibinfo {volume} {106}},\ \bibinfo
  {pages} {013313} (\bibinfo {year} {2022})}\BibitemShut {NoStop}%
\bibitem [{\citenamefont {Kundu}\ and\ \citenamefont {Pati}(2009)}]{Kundu2009}%
  \BibitemOpen
  \bibfield  {author} {\bibinfo {author} {\bibfnamefont {A.}~\bibnamefont
  {Kundu}}\ and\ \bibinfo {author} {\bibfnamefont {S.~K.}\ \bibnamefont
  {Pati}},\ }\href {https://doi.org/10.1209/0295-5075/85/43001} {\bibfield
  {journal} {\bibinfo  {journal} {{EPL} (Europhysics Letters)}\ }\textbf
  {\bibinfo {volume} {85}},\ \bibinfo {pages} {43001} (\bibinfo {year}
  {2009})}\BibitemShut {NoStop}%
\bibitem [{\citenamefont {Di~Liberto}\ \emph {et~al.}(2017)\citenamefont
  {Di~Liberto}, \citenamefont {Recati}, \citenamefont {Carusotto},\ and\
  \citenamefont {Menotti}}]{DiLiberto2017}%
  \BibitemOpen
  \bibfield  {author} {\bibinfo {author} {\bibfnamefont {M.}~\bibnamefont
  {Di~Liberto}}, \bibinfo {author} {\bibfnamefont {A.}~\bibnamefont {Recati}},
  \bibinfo {author} {\bibfnamefont {I.}~\bibnamefont {Carusotto}},\ and\
  \bibinfo {author} {\bibfnamefont {C.}~\bibnamefont {Menotti}},\ }\href
  {https://doi.org/10.1140/epjst/e2016-60388-y} {\bibfield  {journal} {\bibinfo
   {journal} {The European Physical Journal Special Topics}\ }\textbf {\bibinfo
  {volume} {226}},\ \bibinfo {pages} {2751} (\bibinfo {year}
  {2017})}\BibitemShut {NoStop}%
\bibitem [{\citenamefont {Sugimoto}\ \emph {et~al.}(2019)\citenamefont
  {Sugimoto}, \citenamefont {Ejima}, \citenamefont {Lange},\ and\ \citenamefont
  {Fehske}}]{Sugimoto2019}%
  \BibitemOpen
  \bibfield  {author} {\bibinfo {author} {\bibfnamefont {K.}~\bibnamefont
  {Sugimoto}}, \bibinfo {author} {\bibfnamefont {S.}~\bibnamefont {Ejima}},
  \bibinfo {author} {\bibfnamefont {F.}~\bibnamefont {Lange}},\ and\ \bibinfo
  {author} {\bibfnamefont {H.}~\bibnamefont {Fehske}},\ }\href
  {https://doi.org/10.1103/PhysRevA.99.012122} {\bibfield  {journal} {\bibinfo
  {journal} {Phys. Rev. A}\ }\textbf {\bibinfo {volume} {99}},\ \bibinfo
  {pages} {012122} (\bibinfo {year} {2019})}\BibitemShut {NoStop}%
\bibitem [{\citenamefont {Azcona}\ and\ \citenamefont
  {Downing}(2021)}]{Azcona2021}%
  \BibitemOpen
  \bibfield  {author} {\bibinfo {author} {\bibfnamefont {P.~M.}\ \bibnamefont
  {Azcona}}\ and\ \bibinfo {author} {\bibfnamefont {C.~A.}\ \bibnamefont
  {Downing}},\ }\href {https://doi.org/10.1038/s41598-021-91778-z} {\bibfield
  {journal} {\bibinfo  {journal} {Scientific Reports}\ }\textbf {\bibinfo
  {volume} {11}},\ \bibinfo {pages} {12540} (\bibinfo {year}
  {2021})}\BibitemShut {NoStop}%
\bibitem [{\citenamefont {Nakamura}(1999)}]{Nakamura1999}%
  \BibitemOpen
  \bibfield  {author} {\bibinfo {author} {\bibfnamefont {M.}~\bibnamefont
  {Nakamura}},\ }\href {https://doi.org/10.1143/JPSJ.68.3123} {\bibfield
  {journal} {\bibinfo  {journal} {Journal of the Physical Society of Japan}\
  }\textbf {\bibinfo {volume} {68}},\ \bibinfo {pages} {3123} (\bibinfo {year}
  {1999})},\ \Eprint
  {https://arxiv.org/abs/https://doi.org/10.1143/JPSJ.68.3123}
  {https://doi.org/10.1143/JPSJ.68.3123} \BibitemShut {NoStop}%
\bibitem [{\citenamefont {Singh}\ \emph {et~al.}(2014)\citenamefont {Singh},
  \citenamefont {Mishra}, \citenamefont {Pai},\ and\ \citenamefont
  {Das}}]{Singh2014}%
  \BibitemOpen
  \bibfield  {author} {\bibinfo {author} {\bibfnamefont {M.}~\bibnamefont
  {Singh}}, \bibinfo {author} {\bibfnamefont {T.}~\bibnamefont {Mishra}},
  \bibinfo {author} {\bibfnamefont {R.~V.}\ \bibnamefont {Pai}},\ and\ \bibinfo
  {author} {\bibfnamefont {B.~P.}\ \bibnamefont {Das}},\ }\href
  {https://doi.org/10.1103/PhysRevA.90.013625} {\bibfield  {journal} {\bibinfo
  {journal} {Phys. Rev. A}\ }\textbf {\bibinfo {volume} {90}},\ \bibinfo
  {pages} {013625} (\bibinfo {year} {2014})}\BibitemShut {NoStop}%
\end{thebibliography}%

\end{document}